\documentclass[preprint,3p,12pt]{elsarticle}

\usepackage{amssymb}
\usepackage{amsmath}
\usepackage{tabularx,booktabs}
\usepackage{{longtable}}
\usepackage{lscape}
\usepackage{multirow}
\usepackage{multicol}
\usepackage{lineno}
\usepackage{makecell}

\journal{Nuclear Physics B}

\begin{document}

\begin{frontmatter}




\title{Multi-faceted light pollution modelling and its application to the decline of artificial illuminance in France} 

\author[dsl]{Rolf Buhler} 
\author[dsl]{Philippe Deverchère}
\author[dsl]{Christophe Plotard}
\author[dsl]{Sébastien Vauclair}

\affiliation[dsl]{organization={DarkSkyLab},
            addressline={Pascaou}, 
            city={Fontenilles},
            postcode={31470}, 
            state={},
            country={France}}

\ead{info@darkskylab.com}

\begin{abstract}
Artificial Light At Night (ALAN) has been increasing steadily over the past century, particularly during the last decade. This leads to rising light pollution, which is known to have adverse effects on living organisms, including humans. We present a new software package to model light pollution from ground radiance measurements. The software is called \textit{Otus}~3 and incorporates innovative ALAN diffusion models with different atmospheric profiles, cloud covers and urban emission functions. To date, light pollution modelling typically focused on calculating the zenith luminance of the skyglow produced by city lights. In \textit{Otus}~3 we extend this and additionally model the horizontal illuminance on the ground, including the contributions from skyglow and the direct illumination. We applied \textit{Otus}~3 to France using ground radiance data from the Visible Infrared Imaging Radiometer Suite (VIIRS). We calibrated our models using precise sky brightness measurements we obtained over 6 years at 139 different locations and make this dataset publicly available. We produced the first artificial illuminance map for France for the periods of 2013--2018 and 2019--2024. We found that the artificial ground illuminance in the middle of the night decreased by 23\,\% between these two periods, in stark contrast to the global trend. 
\end{abstract}




\end{frontmatter}

\section{Introduction}

As urban areas continue to expand, the excessive and misdirected use of artificial lighting brightens the natural night environment. The consequences of this light pollution are far-reaching, affecting human health \cite{WANG2023121321,Ma2024}, insects \cite{OWENS2020108259}, wildlife \cite{Lao2019,Sanders2020, su11226400, Horton2023}, plants \cite{WangL2025, 10.1093/pnasnexus/pgac046}, energy consumption \cite{barentine_2024_11431447} and our ability to observe the night sky \cite{2022A&ARv..30....1G}. In the recent decade, light pollution has been increasing globally. Satellite measurements observing in the near-infrared to optical bands indicate an average increase of $\approx$2.2\,\% per year worldwide \cite{2017SciA....3E1528K, 2025JQSRT.33509378B}. In the visual band observed by humans, the increase is significantly higher, reaching $\approx$9.6\,\% per year \cite{2023Sci...379..265K}.

In order to address the issue effectively, it is crucial to understand how artificial light spreads and interacts with the environment. This is where light pollution modelling comes into play. It involves using computational techniques and simulations to predict how light is distributed across different landscapes, taking into account atmospheric conditions and sources of artificial light \cite{2000MNRAS.318..641C,2001MNRAS.328..689C,2012MNRAS.427.3337C, 2016SciA....2E0377F, 2020MNRAS.497.2501A}. These models are designed to reveal the spatial and temporal variations of light pollution, helping to develop strategies to minimise its impact.

To date, the main parameter to quantify light pollution is the Night Sky Brightness (NSB). It is historically linked to astronomy and characterizes the visual luminance of the sky around the zenith during clear moonless nights. The NSB level directly affects species that rely on the visibility of stars for orientation \cite{Dreyer2025-hd}. However, for most species the effect of light pollution is better quantified in terms of its total illuminance on the ground \cite{2015MNRAS.446.2895K, 2025MNRAS.542L.154K} (Luminance refers to the brightness perceived by humans from a certain direction, while illuminance refers to the total luminance radiated onto a surface. Note that throughout this paper, we use illuminance referring to an horizontal surface on the ground and luminance at the zenith position.) We will present a novel modelling code, called \textit{Otus}~3, which allows the calculation of zenith luminance and ground illuminance of the skyglow from ground radiance measurements. In addition, we include a model of the direct illuminance, meaning light that directly falls on the ground, without being scattered in the atmosphere. To the best of our knowledge, this makes \textit{Otus}~3 the first software that allows to model the total illumination from artificial light on the ground.

We use \textit{Otus}~3 to model the light pollution in France, where efforts have been done over the past two decades to reduce Artificial Light At Night (ALAN). Light installations are regulated by ministerial orders. They dictate the times at which lighting is permitted and prescribe guidelines for many kinds of lighting installations. In particular, they limit the fraction of light emitted into the sky and the spectral colour of lights \cite{Cerema2020}. In addition, most municipalities in France switch off lighting during the middle of the night \cite{Cerema2025}. We have simulated light pollution parameters for France, to study the effect of these measures. The simulations are based on ground radiance measurements from the Visible Infrared Imaging Radiometer Suite (VIIRS) on-board the Suomi National Polar-orbiting Partnership (Suomi NPP) satellite. This satellite was launched in October 2011 and is a collaborative mission between NASA and NOAA. Its Day/Night Band (DNB) camera provides night-time light images of the entire globe every night.  

The paper is structured as follows: First we discuss the emission and light diffusion models in section \ref{sec:model}. We proceed to describe the VIIRS data, with a particular focus on its observing angle dependency in section \ref{sec:viirs}. In section \ref{sec:calib}, we present how we calibrated our models with a dataset of sky brightness measurements taken over the past six years. We also discuss the expected model accuracy and caveats. In section \ref{sec:france}, we show the results of the model for France.  Finally, we summarize our findings and give an outlook in section \ref{sec:summary}. We use SI units throughout this paper\footnote{Due to its origins in astronomy, NSB is typically reported in units of $mag~arcsec^{-2}$. Due to the small difference between astronomical bands to the photopic sensitivity function the conversion to SI units can vary by $\approx$10\,\%, depending on the spectrum of the observed light \cite{2020MNRAS.493.2429B}. Throughout this work, we use the conversion $NSB[mcd~m^{-2}]~=~1.1~\times 10^{ 8 - 0.4 \times NSB[mag~arcsec^{-2}]}$.}. Results are calculated for photometric quantities, however, only scaling factors would typically change for radiometric quantities.

\section{Model description}
\label{sec:model}

\begin{figure}[ht]
\centering 
\includegraphics[width=\textwidth]{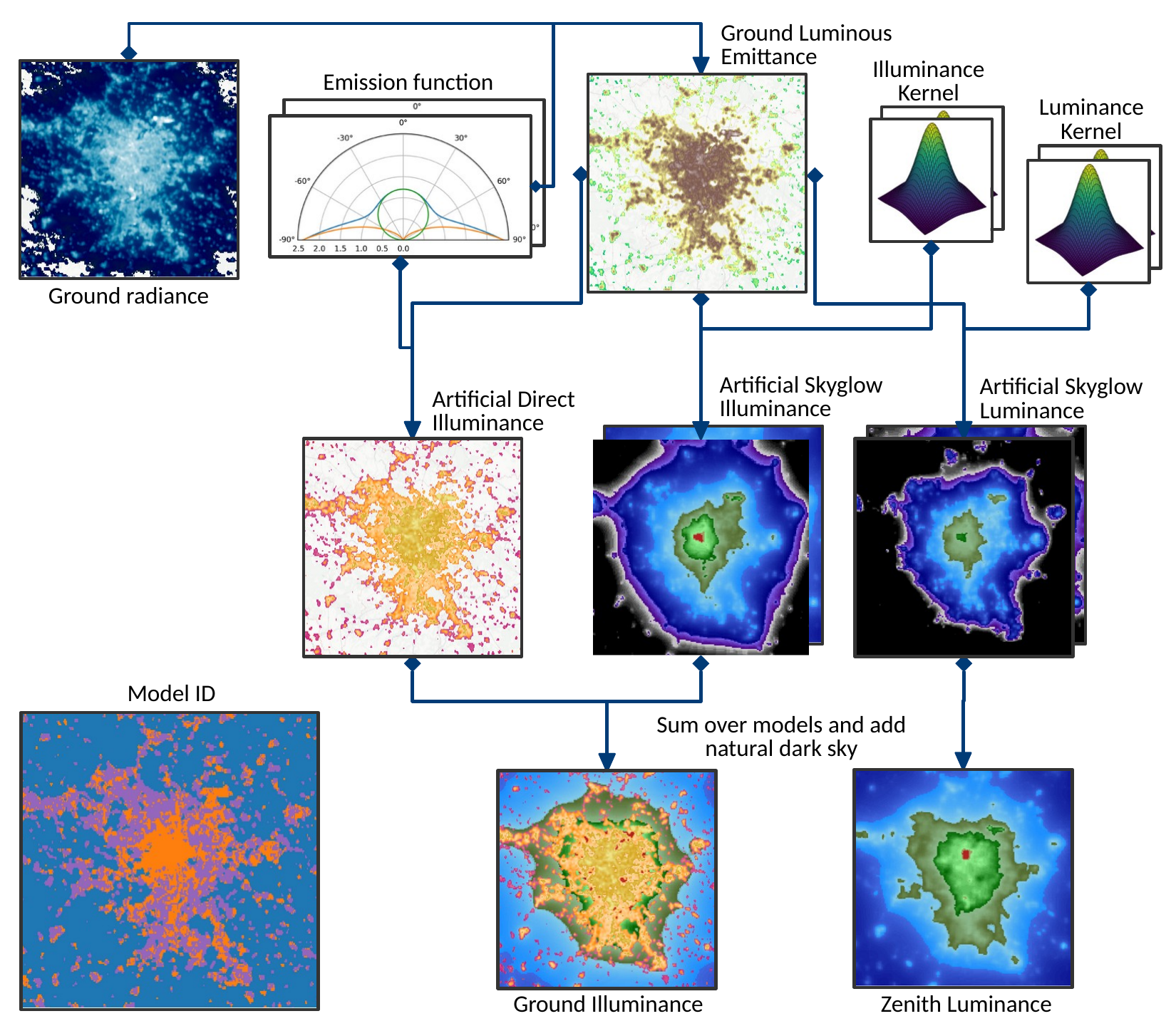}
\caption{Schematic representation of the \textit{Otus}~3 pipeline: ground radiances are transformed to Ground Luminous Emittance (GLE), assuming a pre-defined emission function. The GLE is convolved with diffusion kernels, to calculate the Artificial Skyglow Illuminance and the Artificial Skyglow Luminance. The GLE is also used to calculate the Artificial Direct Illuminance of lamps shining directly on the ground. The natural dark-sky emission from stars and airglow are added to obtain the Total Ground Illuminance and the Total Zenith Luminance. The model parameters, such as atmospheric profile and emission function, can be adjusted for each pixel assigning different Model IDs. Artificial light maps are calculated for each model and summed, which is indicated graphically by the stacking of different maps behind each other. The procedure is repeated for each analysed time step. For more details, see the main text.  \label{fig:schema}}
\end{figure}   

\textit{Otus}~3 simulations are based on an emission model and a subsequent propagation of light in the atmosphere. The emission model is based on observed ground radiance. These components are modular and can be adapted individually; emission and propagation models can be varied spatially and temporally. This is implemented by using a model identification cube, which specifies all model components to be used for any point in space and time. A schematic representation of the structure of \textit{Otus}~3 is shown in figure \ref{fig:schema}. We will describe the different components in the following. 


Following the parameterization introduced by Garstang \citep{1986PASP...98..364G,1989PASP..101..306G}, we characterize the upward radiance from the ground ($RAD$) as a function of the zenith angle ($\theta$) as
\begin{linenomath}
\begin{equation}
RAD(\theta) = k \times GLE \times \frac{(2 G ( 1  - F ) \cos{\theta} + 0.554 F \theta^4)}{2 \pi \cos{\theta}},
\end{equation}
\end{linenomath}
where $GLE$ is the Ground Luminous Emittance of artificial lights, $F$ and $G$ are the angular parameters of the model and $k$ is a scaling parameter. The latter primarily depends on the spectral overlap between the radiance measurement and the emitting lamps. It also depends on surface properties such as the albedo and light masking due to buildings or vegetation. Studies have been done that compare sites of known luminous power with observed VIIRS-DNB radiances \cite{doi:10.1177/1477153513506729, 2014RemS....611915C}. As discussed in appendix \ref{app:radcalib}, adapting these studies to our formalism results in values of $k \approx 260 - 714 \frac{nW / (sr~cm^2)} {lm / m^2}$. In section \ref{sec:calib}, we will discuss how we calibrate $k$ in our analysis.

The Garstang emission function is widely used in light pollution modelling \citep{2000MmSAI..71...93C,2000MNRAS.318..641C}.  Garstang derived it by assuming that the parameter $F$ corresponds to the fraction of light emitted above the horizon and $G$ to the fraction of incoming light reflected on the ground. Reasonable values of ground reflectance (G$\approx$10-20\,\%) and horizontal emission fraction ($F\approx$5-25\,\%) lead to realistic emission functions. While the Garstang's interpretation captures the two primary effects that produce upward light emission, it has been pointed out that in reality its interpretation is more complex, e.g. reflections and blocking of light due to buildings modify the emission function \cite{2000MNRAS.318..641C}. The scaling parameter $k$ accounts for these effects on a phenomenological level.

The Garstang's parametrization provides a relation between the total luminous power emitted by light sources and the radiance emitted in a certain direction of the sky. For an observed radiance $RAD_{obs}$ at a given zenith angle $\theta_{obs}$ it normalizes the emission function within a given model. For given set of Garstang parameters $F$ and $G$, the Ground Luminous Emittance can be calculated as
\begin{linenomath}
\begin{equation}
GLE = \frac{RAD_{obs}}{RAD(k, GLE=1~lm/m^2, \theta_{obs}, F, G)}
\label{eq:garcef}
\end{equation}
\end{linenomath}
and the total Direct Artificial Illuminance (ADI) on the ground as
\begin{linenomath}
\begin{equation}
ADI = GLE \times  ( 1 - F ).
\end{equation}
\end{linenomath}


\begin{figure}[t]
\centering 
\includegraphics[width=\textwidth]{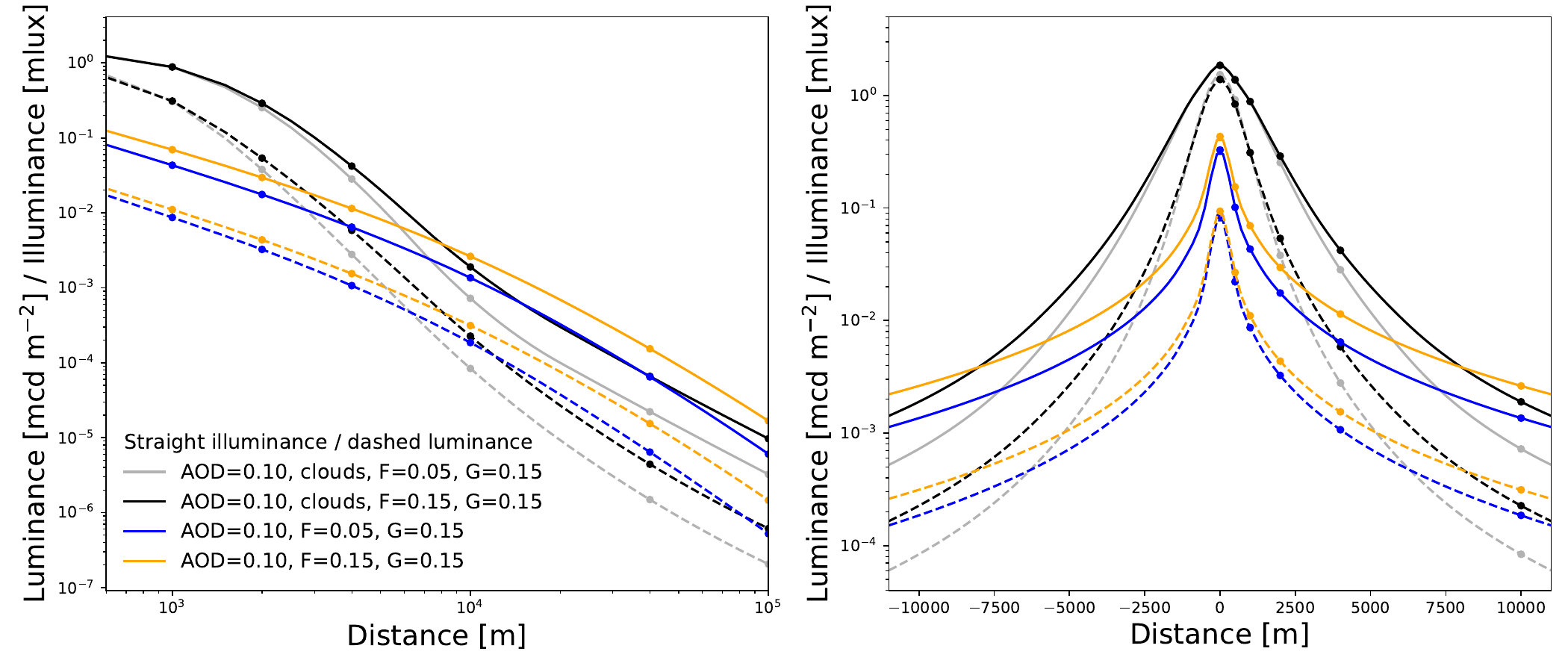}
\caption{Artificial skyglow luminance (dashed lines) and illuminance (straight lines) as a function of distance. Markers indicate distances at which the simulations were done with the \textit{SkyGlow} software. The lines were derived from a 4$^{th}$-order polynomial spline interpolation to these simulations. The latter were done for a square 500m $\times$ 500m region with a  Ground Luminous Emittance of $1~lm/m^2$. The left panel is shown on a logarithmic distance scale, while the right panels is shown in a linear distance scale to visualize the emission around zero. \label{fig:kernels}}
\end{figure}  

Light pollution is not only created by direct illumination, but also by skyglow, which is the light that has been back scattered in the atmosphere. This process depends on the properties of the latter; particularly important are the cloudiness level \cite{2014MNRAS.443.3665K,2016JQSRT.181...11A, 2017NatSR...7.6741J, 2025MNRAS.542L.154K} and the atmospheric aerosol content \cite{1991PASP..103.1109G,2023JEnvM.33517534W, 2023MNRAS.523.2678K}. The backscattered emission is also sensitive to the initial emission function described previously.  We calculate the Artificial Skyglow Luminance (ASL) and Illuminance (ASI) by convolving GLE with kernel functions. The kernel functions vary with the distance of each pixel of a GLE map to the point where the skyglow illumination is calculated. We derived the kernel functions with the help of the \textit{SkyGlow} software \cite{2015MNRAS.446.2895K, 2024MNRAS.533.2356K}. Some examples for different atmospheric conditions are shown in figure \ref{fig:kernels}. The kernel functions can generally be asymmetric, e.g. for a case where the light emission function of a city or the relief around it are asymmetric. The functions and the distances are then replaced by two dimensional arrays. However, in the present study we only considered symmetric kernels. 

\section{Ground radiance measurement}
\label{sec:viirs}

\begin{figure}
\centering 
\includegraphics[width=\textwidth]{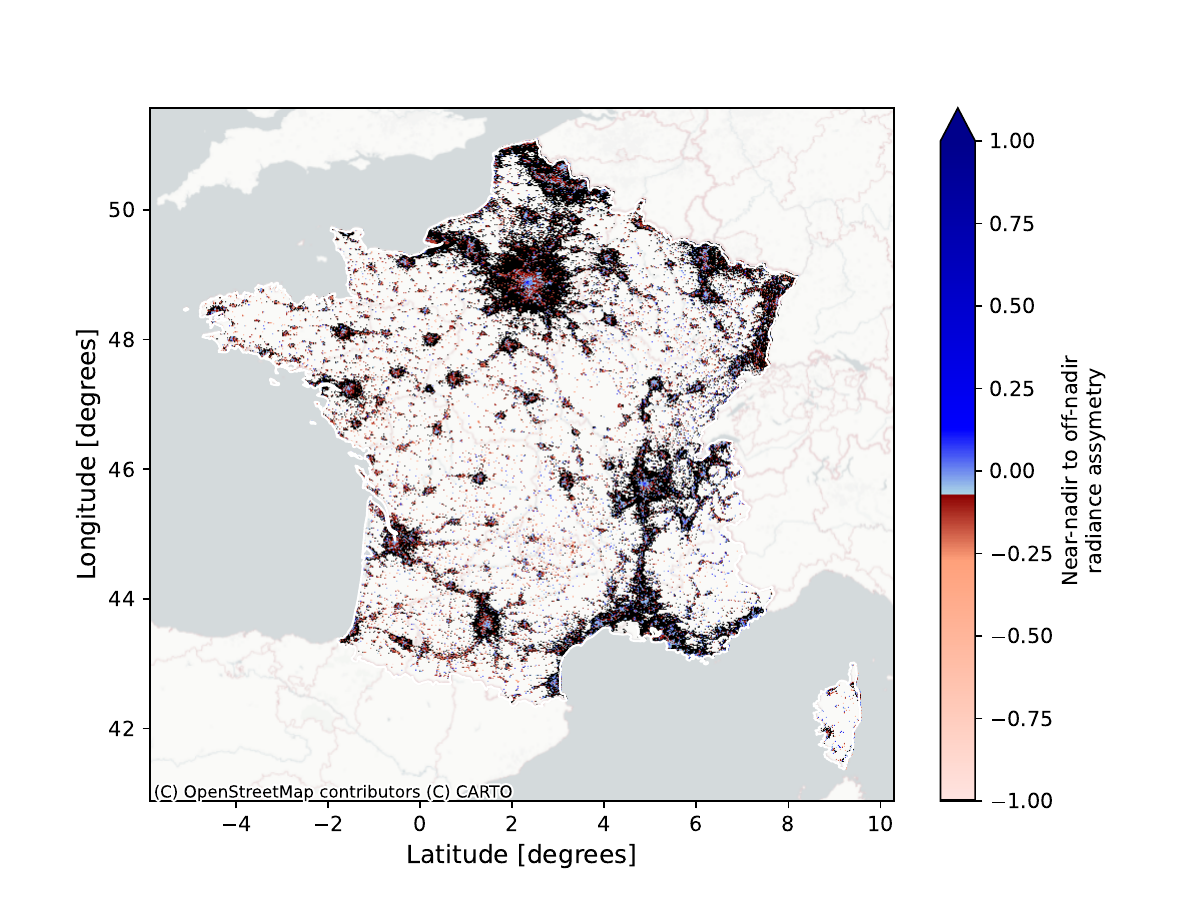}
\caption{This figure shows the asymmetry factor of the ground radiances observed by the VIIRS-DNB instrument between 2019 and 2024. Higher values indicate larger near-nadir radiances compared to the off-nadir radiance. The black areas show regions where ground radiance was removed during the image cleaning, due to the presence of dominant city halo emission (see text for more details).\label{fig:rad_ratio}}
\end{figure}  

The VIIRS-DNB instrument provides ground radiance measurements at night for the entire Earth every day. The exact local time of the measurement can vary between approximately 0:30~am and 2:30~am, depending on the observing angle of the satellite. We refer to this time period, where artificial lighting is usually dimmest, as ``middle of the night''. The DNB camera operates in the visible to near-infrared spectral range of 0.5–0.9~µm and it has a spatial resolution of $\approx$750~m. As a compromise between computation resources and sampling the observing resolution, we applied a 500~m binning to maps throughout this work. We worked in the EPSG:2154 projection, which is well suited for metropolitan France.

We used data processed by NASA's Black Marble calibration pipeline v2.0 \cite{ROMAN2018113, BlackMarbleV2}. This pipeline provides ground radiances, which have already been corrected for atmospheric absorption and scattering.
As satellite observations are often hampered by cloudy weather conditions, monthly- or yearly-averaged composite images are also provided. For the present study, we aimed to obtain images with a low noise that cover a long time period. We therefore split the DNB data into two 6-year time periods, from 2013-2018 and 2019-2024, and averaged their yearly composite images. In order to avoid spurious or ephemerous lights, we only considered radiance which where detected in at least two out of the six years. 

In addition to the standard radiance averaged over all observing angles, the Black Marble pipeline also provides the ground radiance measured in near-nadir (0-20 degrees) and off-nadir (40-60 degrees) observations.  We quantify the difference  between the near-nadir to off-nadir radiances with the asymmetry factor $ASY = \frac{RAD_{near-nadir} - RAD_{off-nadir}}{RAD_{near-nadir} + RAD_{off-nadir}}$. As can be seen in figure \ref{fig:rad_ratio}, the emission asymmetry varies significantly in France and is typically higher in the centres of cities. This is expected due to the masking of horizontal emission by large buildings and has been observed before \cite{2019RSEnv.23311357L}. Later, we will use this information to constrain our emission function.

The VIIRS-DNB radiances do not only contain radiation from the ground, but also faint emission scattered by the atmosphere into the detector. This effect is partially corrected in the image calibration \cite{BlackMarbleV2}. However, a faint "halo emission" around bright cities remains visible in the data. It typically has low ground radiances of 0.5 to 5 nW/(sr~cm$^2$), but covers large areas around bright cities.  It was shown that this halo corresponds to the same skyglow emission that we observe from the ground, when seen from above \cite{sanchez2020}.  It is important to remove it from the data for light pollution parameter simulations, particularly for direct illuminance modelling. Due to its skyglow origin, the halo emission is expected to scale with the Artificial Skyglow Illuminance (ASI) observed in that same pixel. We therefore use the ratio between the observed radiance and the modelled illuminance $RAD / ASI$ to remove halo emission.   The image cleaning has to be applied iteratively, as the cleaning affects the ASI calculation. However, as the halo emission is dominated by the bright city radiances, a good removal is already achieved in one iteration. 

We determined the cut value of $RAD / ASI < 0.35  \frac{nW / (sr~cm^2)} {mlux}$  by visual inspection of different cut values in regions close to cities where a low radiance is expected, such as parks, agricultural areas and the ocean.  We found that the algorithm performed satisfactorily, giving a good trade-off between removing city halo emission and keeping as much radiance from city peripheries as possible. Two examples of this are shown in appendix \ref{app:cleaning}. The areas where ground radiances were removed due to dominant halo emission are shown in black in figure \ref{fig:rad_ratio} . We note however that reality is of course more complex: sub-dominant ground emission can also be present in removed radiance pixels. Also, sub-dominant halo emission is expected to be present in all radiance pixels of a city. However, we consider the presented approach sufficient within the current model accuracy (see section \ref{sec:calib}). 

Finally we note that since 2021 ground radiance data from the Glimmer imager on board the Sustainable Development
Goals Science Satellite 1 (SDGSAT-1) have become available \cite{sdgsat,2024ITGRS..6270572Y}. Compared to the VIIRS-DNB instrument, its images are of higher spatial resolution and provide spectral information. The nightly passage time of the satellite is around 9:30 pm, revisiting every point on Earth approximately every 11 days. These properties make SDGSAT-1 complementary to the VIIRS-DNB and we expect to make use of its data in the future. However, as longer look-back times are not yet available for SDGSAT-1, we did not consider it in the present study.

\section{Model setup and calibration}
\label{sec:calib}


Skyglow models can be calibrated with NSB measurements around zenith done with Sky Quality Meters (SQMs). This is best done using measurements performed at different locations so that the model can be constrained at varying distances from cities \cite{2016SciA....2E0377F}.  Since 2018, we have been conducting a large number of NSB measurement campaigns using a home-developed device called Ninox. Ninox automatically measures the sky brightness at the zenith every minute after sunset. In its current version, Ninox integrates a Unihedron SQM (SQM-LU) which is a widely used photometer when it comes to measure NSB. Its spectral response essentially covers the visible spectrum (roughly from 320 nm to 720 nm in order to match the human vision) and the FWHM of its field of view is 20°. 

Measurements campaigns have been 
conducted in France (metropolitan and oversee regions), Chile and Arizona. They can last from a few nights to a continuous monitoring over several years. These campaigns resulted in a large amount of data with more than 25 million individual NSB measurements across over 400 sites. 


Unfortunately, characterizing night-sky quality through a single NSB value is difficult given the high diversity of meteorological conditions and astronomical configurations (notably the position of the galactic plane) that can influence the measures. In 2022, we proposed \cite{DEVERCHERE2022} a standard indicator called the NSB Dispersion Ratio (NDR) to characterize the luminance level of the sky in both clear sky and cloudy weather conditions. However, this approach is not suitable for reporting a clear-sky NSB indicator that can be reliably used to calibrate the \textit{Otus}~3 model. To address this specific need, we propose here a standard clear-sky NSB indicator, called \textit{Clear Sky Brightness} (CSB). It characterizes the darkest sky observed at a given location in clear-sky conditions. We will discuss the processing steps to derive a CSB in the following.

\subsection{Clear Sky Brightness}

\begin{figure}[t]
\centering
\includegraphics[width=0.8\textwidth]{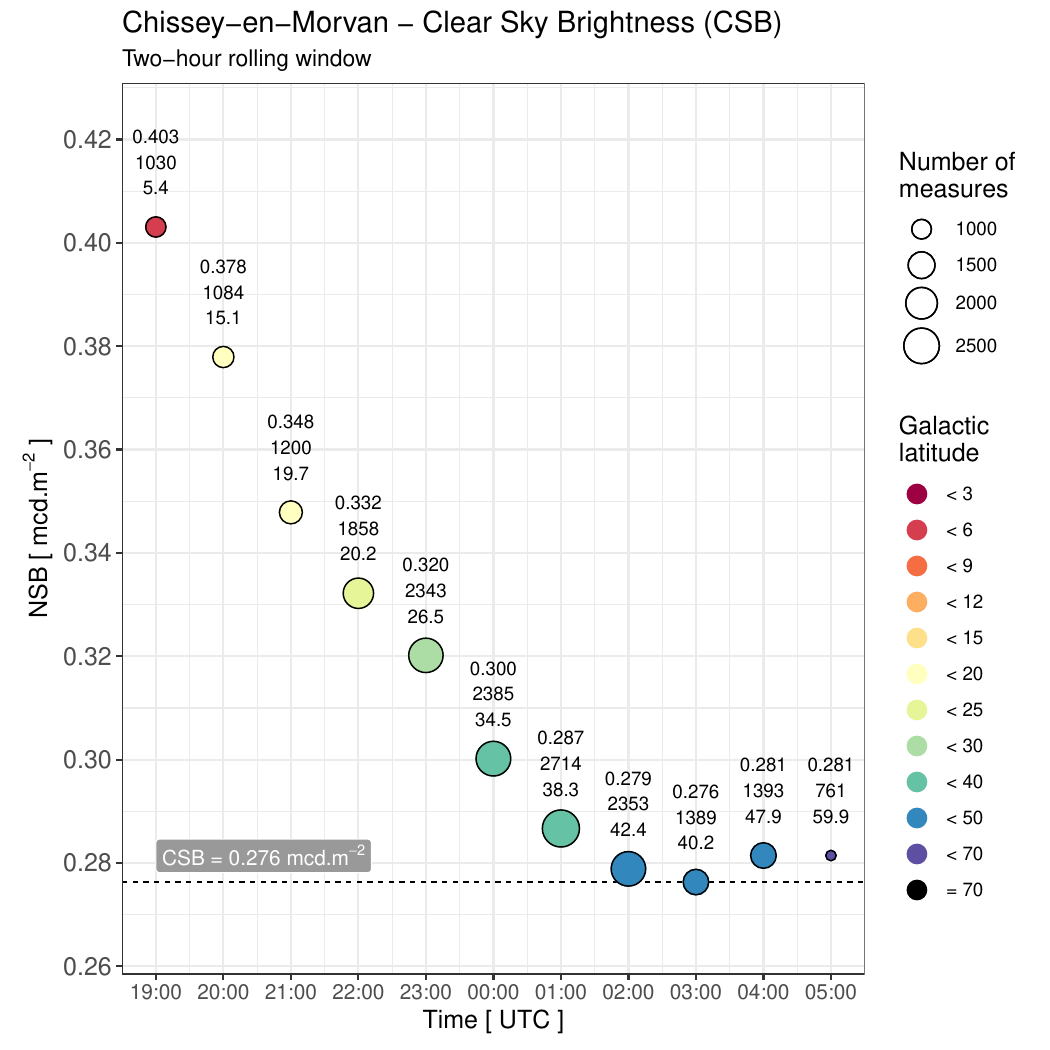}\caption{Clear Sky Brightness plot showing the evolution of the median NSB in two-hour windows at a site within the International Dark Sky Reserve of Morvan, France. In total, 14\,698 clear-sky measures from 127 different nights were used. Above each point the median NSB value, the number of individual measures used to compute it and the average zenithal galactic latitude are provided. The size of each point relates to the number of measures and its colour to the average galactic latitude at zenith.\label{fig:csb_plot}}
\end{figure}

NSB distributions are typically asymmetric with cloudy conditions either above or below the clear sky baseline level depending on the level of light pollution. The CSB characterizes this baseline level. In order to derive it, we need to filter out cloudy nighttime periods and only retain NSB measures from clear-sky nights. For this purpose, a  phenomenological indicator called  \textit{Night Sky Stability} (NSS) has been developed which represents the stability of a series of NSB measures (the cloudier the night, the more chaotic the NSB profile). The method consists of fitting a polynomial to the NSB values of the considered night portion and then calculating the residuals between the measured NSB values and the corresponding polynomial values. A 10$^{th}$-degree polynomial is used for the regression so that the typical variations of the NSB during clear nights can be quite closely fitted. On the other hand, during cloudy nights, the NSB has a high rate of variation which cannot be fit even with a 10$^{th}$-degree polynomial. The NSS  is a unit-less quantity and is obtained by calculating the logarithm of the variance of the residuals $NSS = 7 + log_{10}(Var(NSB - \widehat{NSB}))$, where $\widehat{NSB}$ represents the polynomial value. This formula was derived empirically. Its parameters where chosen such that NSS varies from 0 for an ideal clear night to 10 for highly chaotic NSB curve during a cloudy night for our dataset. We consider two-hour nights portions with an NSS < 2.7 as clear. Examples for this procedure are shown in appendix \ref{app:nss} for a cloudy and a clear night.

After the clear nights have been selected, the CSB can be extracted. Typically one-month, or more, of automated recording is needed to guarantee a few clear nights to perform the measurement. Next, a CSB diagram is created by grouping together all the clear nights of measures (using the daily UTC time) and by calculating the median NSB values in successive two-hour rolling windows exclusively for night portions that are detected as essentially clear (and only for moonless nighttime periods and with the Sun -18° or more below the horizon). There is a shift of one hour between two successive two-hour windows resulting in an overlap in terms of NSB measures used to determine each point in the CSB diagram. This is intentionally done in order to provide a better sensitivity and a one-hour temporal resolution of the indicator.

\begin{table}[t]
    \centering
    \begin{tabular}{lllll}
        \toprule
        \textbf{Model ID}&\textbf{1}&\textbf{2}&\textbf{3}&\textbf{4}\\
        \midrule
        \textbf{Atmosphere}&&&&\\ 
        \midrule
        AOD$_{550}$ &  0.1 & 0.1&  0.1 & 0.1\\
        Aerosol gradient [km$^{-1}$]& 0.65 & 0.65& 0.65 & 0.65  \\
        Molecular scale height [km] &8 & 8 &8 & 8 \\
        Cloud base altitude [km] &  & &1.0 & 1.0 \\
        Cloud fraction &  & & 1.0 & 1.0\\
        \toprule
        \textbf{Emission function}&&&&\\
        \midrule
        F & 0.05&0.15& 0.05&0.15\\
        G & 0.15&0.15& 0.15&0.15\\ 
        k [$\frac{nW / (sr~cm^2)} {lm / m^2}$]&  278 & 371 &  278 & 371 \\
        \toprule
        \textbf{Natural dark sky}&&&&\\
        \midrule
        Zenith luminance [mcd m$^{-2}$]& 0.214 & 0.214 & 0.0 & 0.0 \\
        Ground illuminance [mlux] & 0.7 & 0.7 & 0.0 & 0.0\\ 
        \midrule
    \bottomrule
    \end{tabular}
    \caption{Main atmospheric, emission function and natural sky parameters used in the ALAN simulations of France. For explanations on the individual parameters see the main text and the \textit{SkyGlow} software documentation \cite{SkyGlowWeb}. The emission scaling parameter $k$ and the Natural Zenith Brightness were calibrated with observational data, as discussed in section \ref{sec:modelcalib}.}
    \label{tab:simpars}
\end{table}

Figure \ref{fig:csb_plot} shows the CSB diagram obtained at a good quality rural site in the International Dark Sky Reserve of Morvan, France. The median NSB values have been plotted on a single graph for all two-hour windows with the UTC time on the horizontal axis. The size of the points depends on the number of measurements taken into account in each window (the more measures, the bigger the point), and their color depends on the average value of the absolute zenithal galactic latitude over the same window (a small galactic latitude denotes a strong presence of the Galactic plane at the zenith). Above each point, the median NSB, number of measurements, and average galactic latitude are reported. For the site considered in this example, it can be seen that the best median NSB with a significant number of measurements is obtained for the two-hour window centered on 03:00 UTC, with a value of 0.276~mcd~m\textsuperscript{-2}. This is the CSB, i.e. the characteristic NSB value in clear skies conditions with a minimal contribution of the Galactic plane at the site. We noted that at the beginning of the night, the median NSB values are much higher (representing brighter skies) due to human activities generating light emissions in the night environment (in particular the nearby village did not have its public lighting turned off at 19:00 UTC) as well as a more prominent presence of the Milky Way (red and yellow points).

\subsection{Model parameters and scaling}
\label{sec:modelcalib}

To setup our simulations we chose two different emission functions, based on the radiance asymmetry measured in the respective six-year composites. For the time period of 2019-2024 the asymmetry factor for France is shown in in figure \ref{fig:rad_ratio}.  The first model, for areas with $ASY > -0.07$, has low horizontal emission ($F=0.05$, with Model ID~=~1). It typically characterises city centres or other areas with significant blocking of horizontal light. The second model, for areas with $ASY < -0.07$, has a high horizontal emission ($F=0.15$, with Model ID~=~2). It typically applies to towns, suburbs or rural areas. The ground reflectivity was set to $G=0.15$ for both models. The atmospheric parameters of the \textit{SkyGlow} simulations were set to typical values for central Europe, in particular, the Aerosol Optical Depth was set to 0.1 \cite{DiAntonio2023}. We simulated light diffusion kernels for clear and low-cloud atmospheric conditions. A summary of the main simulation parameters is listed in table \ref{tab:simpars}.

To calibrate our two models, we performed a maximum likelihood fit of the simulated zenith luminances to the CSB measurements. For this, we chose 139 sites in Metropolitan France which had enough NSB measurements to derive a CSB and were well covered by VIIRS observations. The measured values for all sites are publicly available  \cite{philippe_deverchere_2025_17098451}. The simulations were performed on the average monthly-radiance composites covering the same time period as the one of the CSB data taking. Besides the scaling parameter $k$ of each model we varied the average natural-sky zenith luminance (NZL) in the fit. We note, that in order to convert RAD to GLE for composite images with equation \ref{eq:garcef}, an integration over all observations angles needs to be performed. However, the observing angle distribution is not provided by for the Black Marble composites. We therefore assumed an average observing angle of 30 degrees. We verified that using the mean nadir angle leads to differences of only a few percent compared to doing a full angle integration, well below our model accuracies discussed in section \ref{subsec:caveats}.


\begin{figure}[t]
\centering
\includegraphics[width=0.8\textwidth]{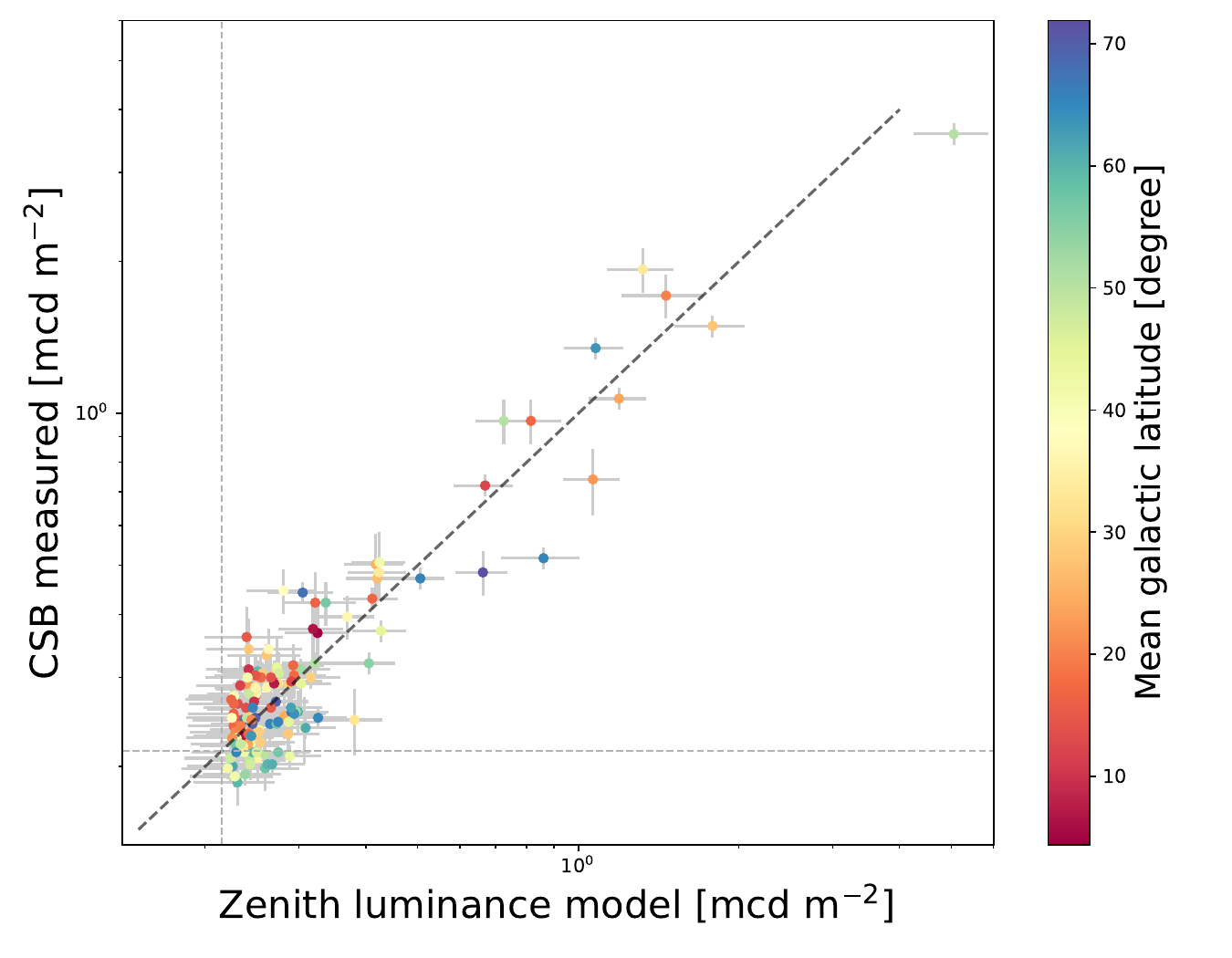}
\caption{Measured CSB values as a function of simulated zenith brightness. The black dashed line shows the points where both are equal. The gray dashed lines show the best-fit natural sky zenith luminance of $0.214~mcd~m^{-2}$. The colour shows the mean galactic latitude of the Milky way during the CSB data taking. }
\label{fig:calib}
\end{figure} 

A comparison of the measured CSB and simulated zenith brightness values is shown in figure \ref{fig:calib}, together with the best fit curve. One can see that a satisfactory agreement is achieved. The best fit parameters are listed in table \ref{tab:simpars}. The average $NZL~=~0.214~mcd~m^{-2}$ is in line with expectations \cite{2021AJ....162...25A}. We emphasize that the NZL is measured in the fitting procedure and is not assumed ad hoc, as is often done in ALAN simulations. It therefore represents a realistic average value for France during the time period we are considering. The average galactic latitude at zenith during the CSB measurement is shown in colors in figure \ref{fig:calib}. One can see a clear dependence of the CSB on the Milky way position remains. To further improve the accuracy of the simulations at these low CSB levels a dynamic model of the Milky way emission in the sky is needed, which is beyond the scope of this paper.

\subsection{Model accuracy and caveats}
\label{subsec:caveats}

\begin{table}[t]
\centering
    \begin{tabular}{lllll}
        \toprule
         & \textbf{Zenith Luminance} & \multicolumn{3}{c}{\textbf{Ground Illuminance}} \\
        & & Skyglow & Direct & Natural\\
    \midrule
    Natural moonless clear sky     & 0.04 mcd m$^{-2}$&  &  & 0.2 mlux\\
    Model atmosphere \& emission     & 18\,\% & 18\,\% & 49\,\% & \\
    VIIRS-DNB sensitivity limit    &  &  & 25 mlux & \\
    \bottomrule
    \end{tabular}
    \caption{Summary of the estimated systematic uncertainties of the light pollution parameters.}
    \label{tab:errors}
\end{table}

Simulated light pollution parameters have significant uncertainties, as they depend on assumptions on the atmospheric and emission function properties. We have summarized the model uncertainties we estimate for our simulations in table \ref{tab:errors} and will discuss them in the following. Note that statistical measurement errors of the radiance measurement also lead to uncertainties in light pollution parameters. In the case of the current study however, we averaged radiance data over many years. The radiance measurement errors are therefore smaller than the modelling uncertainties. 

We estimate the uncertainty on zenith luminance based on the level of agreement between the measured and simulated values shown in figure \ref{fig:calib}. After considering radiance and CSB measurement  errors, a significant scatter remains in this correlation. As mentioned, at low zenith luminance this is partly due to variations in the natural sky. We estimate the uncertainty in this component to be $\Delta NZL = 0.04~mcd~m^{-2}$. We attribute all remaining scatter to uncertainties in the ASL simulation itself. To quantify it, we increased the relative errors on ASL, adding it in quadrature to the other errors until we obtained a reduced chi squared of one between the measured and simulated luminance values. The resulting uncertainty is $\Delta ASL / ASL=18\,\%$.  

For ground illuminance very few measurements exist so far. We therefore cannot derive its uncertainty directly from measurements. However, from a modelling stand point there is nothing special about simulating the skyglow emission from zenith compared to all other directions that are integrated to obtain ASI. We therefore assume an uncertainty of $\Delta ASI / ASI \approx 18\,\%$, as for ASL. We also do not have a measurement of the Natural Ground Illuminance (NGI) during moonless dark nights at this point. We therefore took a value of $NGI \approx 0.7~mlux$ from the literature, obtained by integrating the luminance over all different sky directions \cite{2012MNRAS.427.3337C, 2015MNRAS.446.2895K}. We estimate the uncertainty on the latter as $\Delta NGI \approx 0.2~mlux$, a similar relative error as for $NZL$.

For direct illuminance the uncertainties are expected to be larger than for the skyglow components. The reason is that direct illuminance is calculated locally in each radiance pixel, model assumptions therefore do not average out over many pixels. While the atmospheric model does not play a role, the calculation of direct illuminance is sensitive on the emission function. We estimated the uncertainty resulting from this by varying the emission function parameters G, F, $\theta_{obs}$ and k within reasonable ranges and calculating its impact on ADI ([0.1--0.2], [0.05--0.25], [25--35 deg], [250--450~$\frac{nW / (sr~cm^2)} {lm / m^2}$], respectively). The distribution resulting from all possible parameter combinations on these parameters has a relative width of $\Delta ADI / ADI = 49\,\%$, which we take as an estimate of the uncertainty on ADI.

In addition to the model uncertainty, ADI measurements are limited by the minimum radiance measured by VIIRS of $RAD_{min} \approx 0.5~nW~sr^{-1}~cm^{-2}$. (Note, that this sensitivity limitation does not affect to the long range skyglow emission as strongly, as skyglow emission is dominated by brighter city radiances).  The minimum detection threshold translates into a minimum $ADI \approx 25~mlux$, corresponding to a minimum $GLE \approx 0.03~lm~m^{-2}$. In practice, this means that at minimum two street lamps, with a power of $\approx 8000~lm$ each, is typically required over the area of $750~m~\times~750~m$ for their emission to be detectable by VIIRS. Inspection of the ADI maps shows that pixels below the detection threshold usually correspond to rural areas, e.g. fields or forests. No significant emission is usually expected in such areas in the middle of the night; lights typically cluster in towns, which are in their large majority detected by VIIRS. We therefore expect that only a small fraction of pixels have significant emission below the detection threshold ($\lesssim $1\,\%). Nevertheless, it is important to keep this caveat in mind when using direct illuminance maps based on VIIRS data.

{Finally we note that we have calibrated the model using Ninox data which was taken from 2018 onward. At this point, there is not enough data available to perform a time dependent calibration for the entire time period considered in this work. While the time evolution of light pollution parameters is primarily affected by changes in the observed radiance, more subtle changes can come from changing emission functions or spectra of light installations. Changing emission functions are considered in our analysis by using the asymmetry factor of the six-year composites for the corresponding time period to derive the model maps. However, spectral changes in the installed lighting could in principle lead unaccounted variations. In particular, it is well known that the installation of LEDs can lead to blue radiance bellow 500 nm, which cannot be detected by the VIIRS-DNB instrument and is therefore not reflected in the radiance time evolution \cite{2016SciA....2E0377F,doi:10.1126/sciadv.abl6891}. However, in our case we do not expect this to be an important factor for two reasons: (1) We use photopic quantities throughout this work. The photopic sensitivity curve is low for emission below 500 nm and is well covered by the VIIRS sensitivity band. (2) Public lighting in France must have a colour temperature below 3000~K \cite{Cerema2020}. We therefore do not expect a strong blue radiance contribution from new LED installations for France during the middle of the night. }

\section{Light pollution in France}
\label{sec:france}

\begin{table}[t] 
\centering
\caption{Colour scale to visualize ground illuminance and zenith luminance. It extends the Bortle scale, which subdivides the quality of the dark night sky into different levels (starting at one for the darkest sky; for more details see the main text). The colour shading within each level is done on a logarithmic scale. \label{tab:colour}}
\begin{tabular}[t]{lcllll}
\toprule
\makecell{\textbf{Ratio to} \\  \textbf{natural} \\  \textbf{dark sky}} & \makecell{\textbf{Extended} \\ \textbf{Bortle} \\ \textbf{scale}} &\makecell{\textbf{luminance} \\ \textbf{[mcd m$^{-2}$]}}&\makecell{\textbf{luminance} \\ \textbf{[mag arcsec-2]}}& \makecell{\textbf{illuminance} \\ \textbf{[mlux]}} & \textbf{Color} \tabularnewline
\midrule

3981 -- 10000&11&694.1 -- 1743&13.00 -- 12.00 &2508 -- 6300& \includegraphics[width=0.07\linewidth]{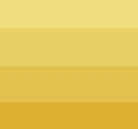} \tabularnewline

1000 -- 3981&10&174.3 -- 694.1&14.50 -- 13.00&630.0 -- 2508& \includegraphics[width=0.07\linewidth]{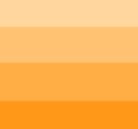} \tabularnewline

158.5 -- 1000&9&27.63 -- 174.3&16.50 -- 14.50&99.85 -- 630.0&\includegraphics[width=0.07\linewidth]{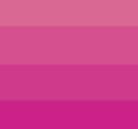} \tabularnewline

39.81 -- 158.5&8&6.941 -- 27.63&18.00 -- 16.50&25.08 -- 99.85& \includegraphics[width=0.07\linewidth]{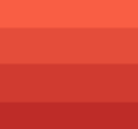} \tabularnewline

25.12 -- 39.81&7&4.379 -- 6.941&18.50 -- 18.00&15.82 -- 25.08& \includegraphics[width=0.07\linewidth]{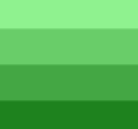} \tabularnewline

12.59 -- 25.12&6&2.195 -- 4.379&19.25 -- 18.50&7.93 -- 15.82& \includegraphics[width=0.07\linewidth]{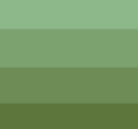} \tabularnewline

4.79 -- 12.59&5&0.834 -- 2.195&20.30 -- 19.25&3.02 -- 7.93& \includegraphics[width=0.07\linewidth]{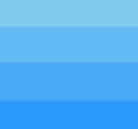} \tabularnewline

1.91 -- 4.79&4&0.332 -- 0.834&21.30 -- 20.30&1.20 -- 3.02& \includegraphics[width=0.07\linewidth]{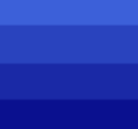} \tabularnewline

1.45 -- 1.91&3&0.252 -- 0.332&21.60 -- 21.30&0.91 -- 1.20& \includegraphics[width=0.07\linewidth]{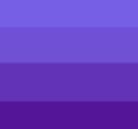} \tabularnewline

1.26 -- 1.45& 2	& 0.219  -- 0.252 &21.75 -- 21.60&0.79 -- 0.91& \includegraphics[width=0.07\linewidth]{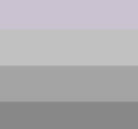} \tabularnewline
1.00 -- 1.26& 1 & 0.174 -- 0.219 &22 -- 21.75&0.63 -- 0.79& \includegraphics[width=0.07\linewidth]{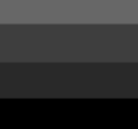} \tabularnewline

\bottomrule
\end{tabular}
\end{table}

\begin{figure}
\centering
\includegraphics[width=\textwidth]{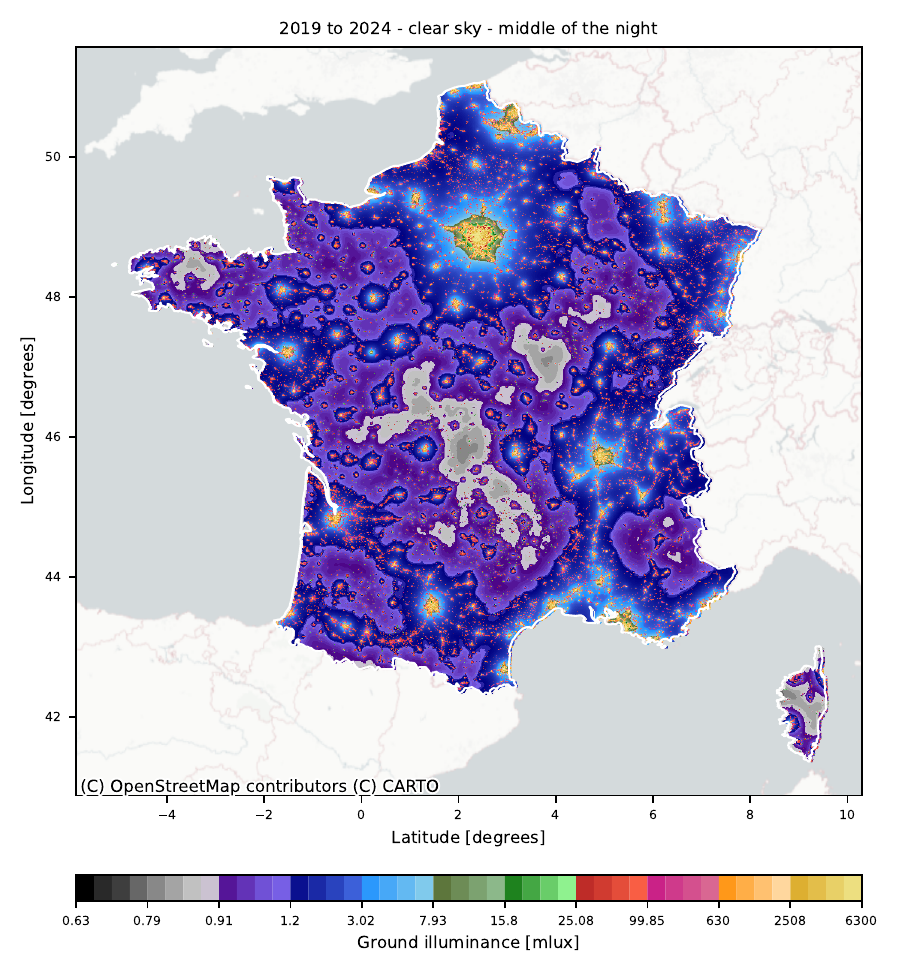}
\caption{Simulated ground illuminance for France in the middle of the night for clear sky conditions. The simulations are based on VIIRS ground radiance measurements between 2019 and 2024.  The model parameters are listed in table \ref{tab:simpars}. The ground illuminance is composed of direct illuminance and the diffuse illuminance from sky glow and the natural dark sky. These components are shown separately in appendix \ref{app:maps}. The illuminance values represent averages over a pixel size of $500 m \times 500 m$. The estimated accuracy of the simulation is listed in table \ref{tab:errors}. \label{fig:tgi}}
\end{figure}

\begin{figure}
\centering
\includegraphics[width=\textwidth]{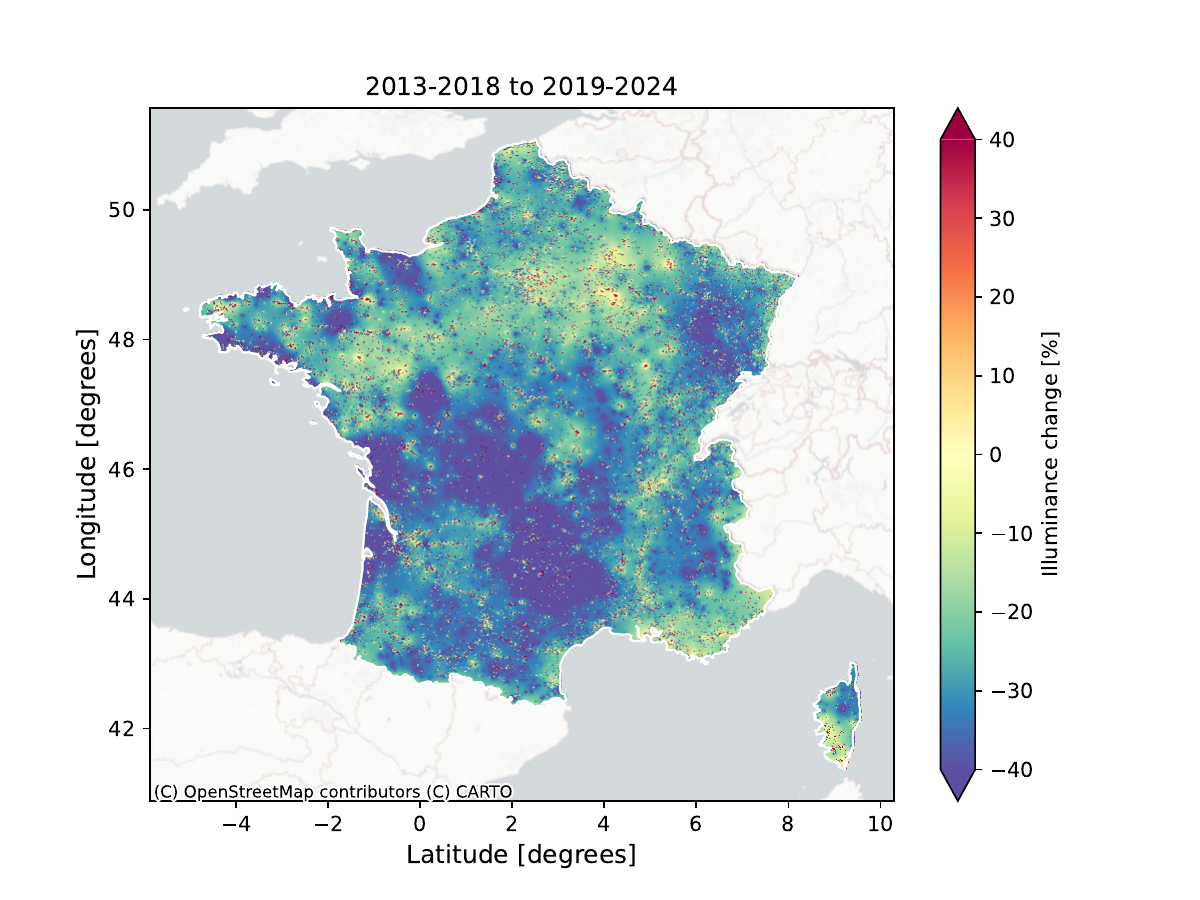}
\caption{Relative change of the artifical ground illuminance in the middle of the night for metropolitan France between 2013--2018 and 2019--2024. \label{fig:evolution}}
\end{figure} 

We have simulated light pollution  parameters for France between 2013--2018 and 2019--2024 using the setup described in section \ref{sec:modelcalib}. To visualize our results, we created a dedicated colour scale, which has colour transitions at the Bortle scale thresholds. The latter divides the quality of the dark sky into nine categories, taking the zenith luminance of a dark natural sky as a reference baseline \cite{Bortle2001}.  We extended this approach by applying the same relative ratios to the natural sky to ground illuminance, assuming value of $NGI=0.63~mlux$, corresponding to the lowest values we found in the literature \cite{2015MNRAS.446.2895K}. The inclusion of direct illuminance to our analysis leads to significant higher artificial light ratios compared to the natural sky. We therefore extended the Bortle scale by two levels at the high-illuminance end. The different thresholds and the corresponding colours are listed in table \ref{tab:colour}.

The simulation result for zenith luminance is shown in appendix \ref{app:maps}, for comparison with other works.  Here we focus on the results for illuminance, which is the main novelty of \textit{Otus}~3. The total ground illuminance for the time period of 2019--2024 is shown in figure \ref{fig:tgi}. The latter is composed of direct illuminance and diffuse sky illuminance, produced by skyglow and natural-sky emission. These components are also shown separately in appendix \ref{app:maps}. One can see, that the diffuse illuminance is systematically lower than the direct illuminance. Outside of the centres of large cities, it is always below the detection threshold of $ADI \approx 25~mlux$. For this reason, the diffuse emission is predominantly shown in colours from gray to green. One can see that it varies smoothly over long distances. In contrast, the direct illuminance is predominantly shown in red to yellow colours. It is concentrated in the urban areas and has a much finer structure, often following rivers or mayor highways. The darkest night sky in France is found in the Massif Central, a mountainous region which is relatively sparsely populated. Dark night skies are also found in Brittany, Corsica and Nièvre.

We emphasize, that the illuminance values in figure \ref{fig:tgi} represent average values over the pixel size of $500 m \times 500 m$. Illuminance is expected to vary strongly within each pixel, as the ground albedo, lighting properties and masking vary significantly on smaller spatial scales. Illuminance maps based on VIIRS-DNB data are therefore primarily useful to highlight areas where light sources are concentrated and not for a lighting analysis on the typical spatial scale of buildings or streets.

The evolution of the artificial ground illuminance, produced by skyglow and direct illumination, is shown in figure \ref{fig:evolution}. One can see a general trend of reduction across the territory, with localized exceptions where new infrastructures were built. The total decrease of artificial ground illuminance between 2013--2018 and 2019--2024 is 23\,\%. The reduction is most pronounced in rural areas around small towns or cities. This is where the switching off of public lighting is most commonly practised.

\begin{figure}
\centering
\includegraphics[width=\textwidth]{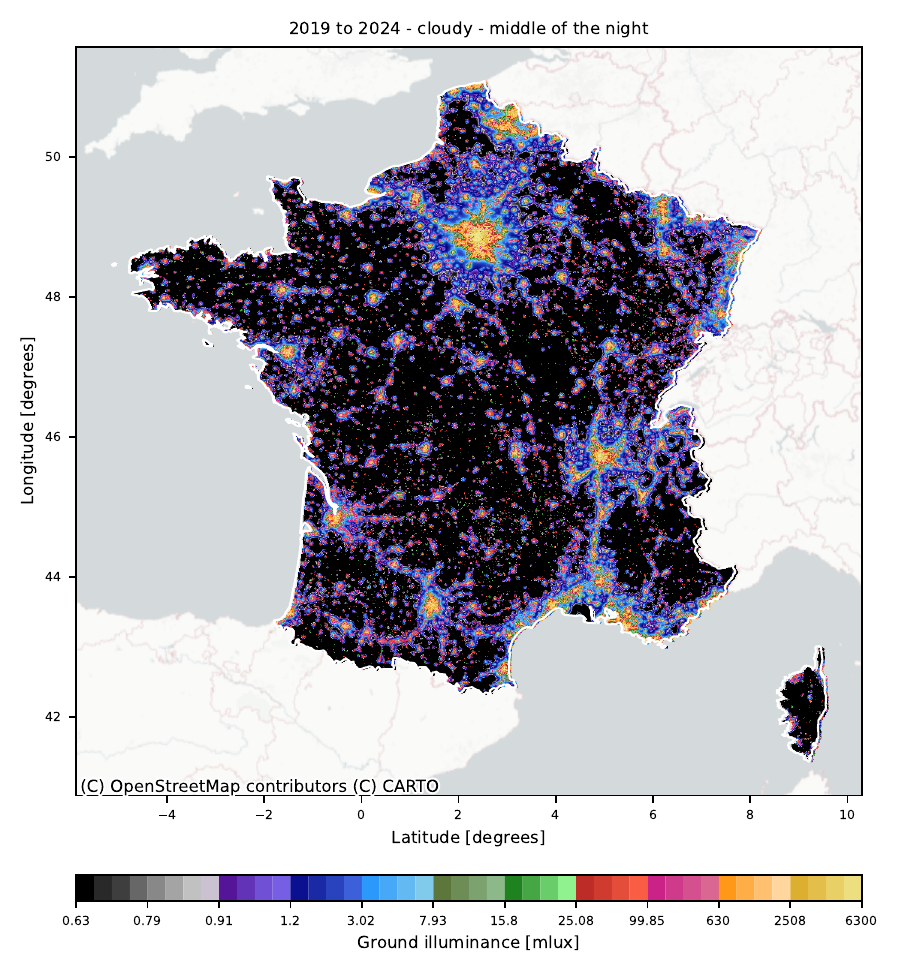}
\caption{Same as figure \ref{fig:tgi}, but for full low-cloud coverage. The simulation parameters are listed in table \ref{tab:simpars}. }
\label{fig:tgicloud}
\end{figure}

A large fraction of the time the sky is covered by clouds. In France, this fraction can vary from $\approx 25\,\%$ in summer to $\approx 60\,\%$ in winter \cite{Baray2019}. Therefore, it is important to consider the light pollution during cloudy conditions. As mentioned, cloud properties vary strongly in time and are not homogenous over the entire territory. To illustrate the maximum effect that cloud coverage can have, we chose an extreme scenario of low clouds (1~km) with 100\,\% coverage over the entire territory. Analogously to the clear-sky simulations, we assigned model IDs 3 and 4 listed in table \ref{tab:simpars}. The result is shown in figure \ref{fig:tgicloud}. As expected, clouds darken the sky far away from urban areas, as they block the propagation of ALAN to large distances. One can also see that rural areas that already had a dark night sky darken even further. The reason is that, besides the skyglow emission, clouds also block the natural sky produced by stars and air glow. The opposite effect is seen within and around urban areas, where reflection of city lights by clouds lead to an increase of Artificial Skyglow Illuminance of up to an order of magnitude.

\section{Summary and outlook}
\label{sec:summary}

We have described a new software, called \textit{Otus}~3, that allows the simulation of light pollution parameters from ground radiance measurements. In addition to zenith luminance, the code calculates the total illuminance on the ground. The latter includes contributions for skyglow and direct ground illumination. We support the view that total ground illuminance shall be used in the future to quantify light pollution, in addition to the zenith luminance used in the past. For many living organisms ground illuminance relates directly to the effects ALAN has on them \cite{2015MNRAS.446.2895K}.


We applied \textit{Otus}~3 to metropolitan France using ground radiance measurements from the VIIRS-DNB instrument. We simulated ground illuminance maps for the time periods of 2013--2018 and 2019--2024 for clear and cloudy sky conditions. We calibrated our simulations with zenith luminance measurements we performed at 139 different sites over the past six years.  We believe this to be an unique dataset in terms of quality and made the results publicly available \cite{philippe_deverchere_2025_17098451}. We hope that it can be useful to the community for future ALAN studies. We will continue our measurement campaigns and plan to include data from other devices to our calibration, e.g. from the TESS photometers \cite{TESSManual}. We also plan to add high-sensitivity lux meters to our Ninox measuring devices. This will allow us to build up a dataset of ground illuminance measurements to cross-calibrate our simulations in an independent way.

We have shown that on average the artificial ground illuminance decreased by 23\,\% in metropolitan France between 2013--2018 and 2019--2024. This reduction is a great success, particularly considering the steep global increase in light pollution over the past decade. It shows that the practice of switching off lights at night, which is widely applied in France, is a powerful tool to limit ALAN emission. However, we emphasize that the presented results apply to the middle of the night only, when human activity is minimal. The ALAN reduction (if any) is expected to be much smaller at the beginning and end of the night when public lighting is not switched off. We plan to study light pollution at both ends of the night in future works.

\section*{Acknowledgement}

We would like to thank all the representatives of the national parks, regional natural parks and municipalities, as well as the individual contributors who made the long Ninox measurement sessions possible. We would also like to thank the developers of the SkyGlow software for making their tool freely available and for their prompt and effective support.

\section*{Abbreviations}{
The following abbreviations are used in this manuscript:
\\
\noindent 
\begin{tabular}[t]{@{}>{\raggedright\arraybackslash} llp{4.5cm}l}
\toprule
\textbf{Name} & \textbf{Acronym} &\textbf{Description}& \textbf{Unit}\\ 
\midrule
Artificial Direct Illuminance & ADI &  Illuminance on the ground directly from artificial lights & mlux \\
Artificial Light At Night& ALAN & Light emitted by human devices at night&\\
Artificial Skyglow Illuminance & ASI &  Illuminance from skyglow on the ground &mlux \\
Artificial Skyglow Luminance & ASL &  Luminance from skyglow at zenith & mcd m$^{-2}$ \\
Clear Sky Brightness & CSB & Characteristic Night Sky Brightness for a given location & mcd m$^{-2}$ \\
Diffuse illuminance  &  &  Illuminance produced by skyglow and the natural dark sky & mlux  \\
Ground illuminance & &  Total illuminance on the ground, including artificial and natural light & mlux \\
Ground Luminous Emittance &GLE &  Luminous flux emitted by light sources on the ground & lm m$^{-2}$\\
Ground Radiance & RAD & Radiance emitted from the Earth surface into the atmosphere & nW sr$^{-1}$ cm$^{-2}$\\
Natural Ground Illuminance & NGI &  Illuminance of the dark moonless sky & mlux \\
Natural Zenith Luminance  & NZL &  Luminance of the dark moonless sky at zenith & mcd m$^{-2}$  \\
Night Sky Brightness & NSB &  Luminance averaged 10 degrees around the zenith & mcd m$^{-2}$\\
Zenith luminance  & &  Total luminance at zenith, including artificial and natural light & mcd m$^{-2}$  \\
\bottomrule
\end{tabular}
}

\appendix

\section{Radiance emission function calibration}
\label{app:radcalib}
A comparison of observed radiances from different sites has been performed to assess the calibration and stability of the VIIRS-DNB camera in reference \citep{2014RemS....611915C}. The authors selected shipping vessels, the Geneva city center and the San Matteo and Incheon bridge and compared the observed radiance to the one expected from the luminous power of these sites. In our context the study of the San Matteo Bridge is most useful, as the luminous power and atmospheric conditions were relatively well known. 
The radiance at this site was measured to $RAD_{obs} = 3.28~nW / (sr~cm^2 )$ at an observing angle of $\theta_{obs}\approx 32$~deg. White LED lamps with an electric power consumption of 310 Watt were installed at the bridge. The authors estimate a radiant power efficiency of 30\,\%, out of which 66\,\% are expected to be within the DNB band. If we assume a further conversion factor of 30\,\% into the visible range, each lamp is expected to have a luminous power of  $\approx $ 20000~lm. With a mean distance between lamps of 54.86~m this results in GLE=1.16 lm~m$^{-2}$ within each VIIRS-DNB pixel. For the emission function the authors consider only ground reflections, with an effective reflectivity of G=0.04, which takes into account that a significant fraction of light is absorbed in the water around the bridge. Inserting these parameters in equation \ref{eq:garcef}, results in $k \approx 260 \frac{nW / (sr~cm^2)} {lm / m^2}$ .

Another study was carried out in the city of Flagstaff, USA \cite{doi:10.1177/1477153513506729}. The total luminous power of the city was estimated to $\approx$158 Mlm. The integrated radiance, over the entire area of the city was 42725 Watt/sr. The city emission function was assumed to be parametrized by a Garstang emission function with G=0.15 and F=0.10. Assuming an observing angle of $\theta_{obs}\approx 30$~deg, this results in $k = 714 \frac{nW / (sr~cm^2)} {lm / m^2}$.

\section{Additional figures on radiance image cleaning}
\label{app:cleaning}
\begin{figure}
\centering
\includegraphics[width=0.9\textwidth]{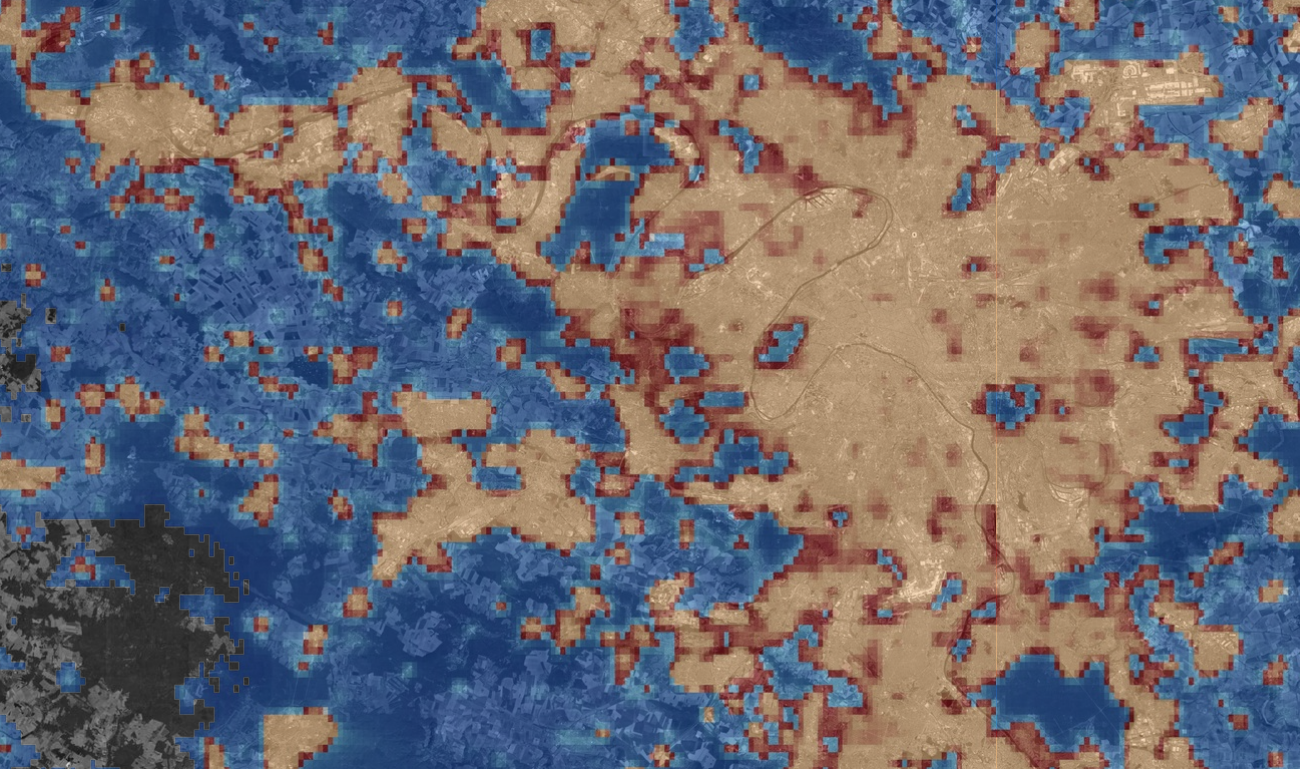}
\includegraphics[width=0.9\textwidth]{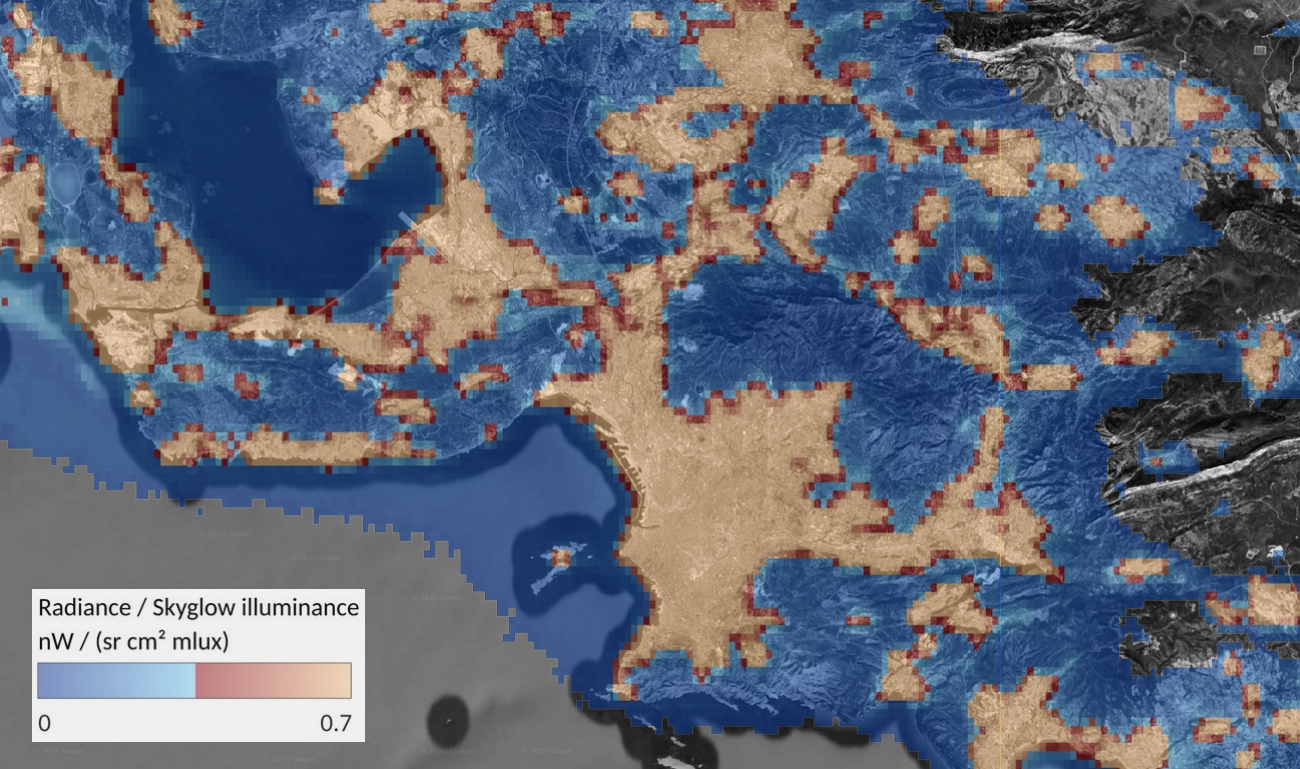}
\caption{These figures show the ratio of the radiance measured by VIIRS between 2019 and 2024 to the artificial skyglow illuminance for Paris (top panel) and Marseille (bottom panel). The colour scale was chosen to illustrate the image cleaning discussed in section \ref{sec:viirs}: areas where the ground radiance was kept are shown in reddish colours and zones where radiance was removed in blueish colours (credit: NASA, NOAA, Google).}
\end{figure}  

\section{Additional figures on Night Sky Stability derivation}
\label{app:nss}
\begin{figure}
\centering
\includegraphics[width=0.7\textwidth]{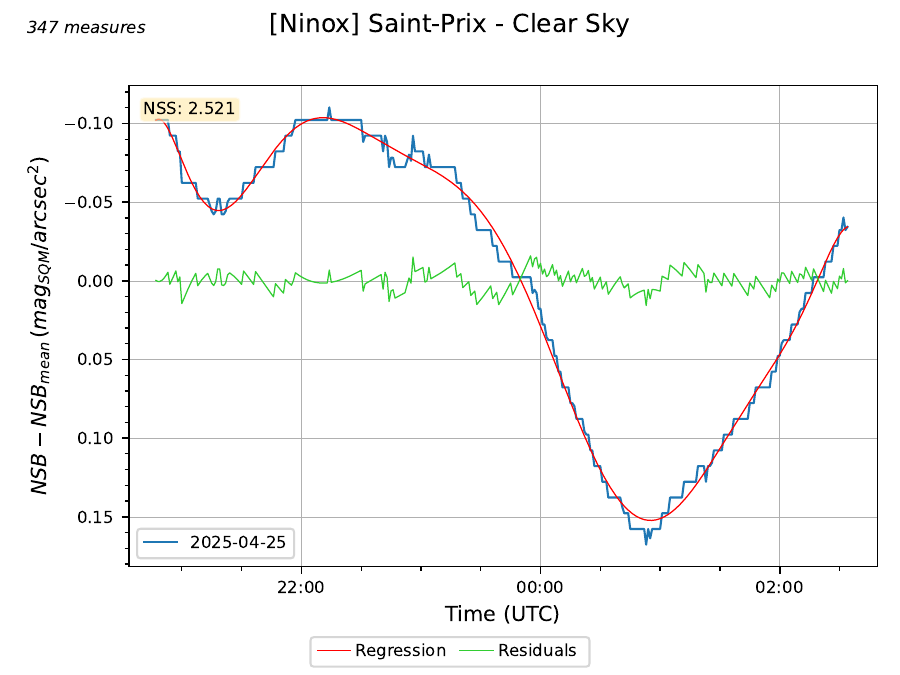}
\includegraphics[width=0.7\textwidth]{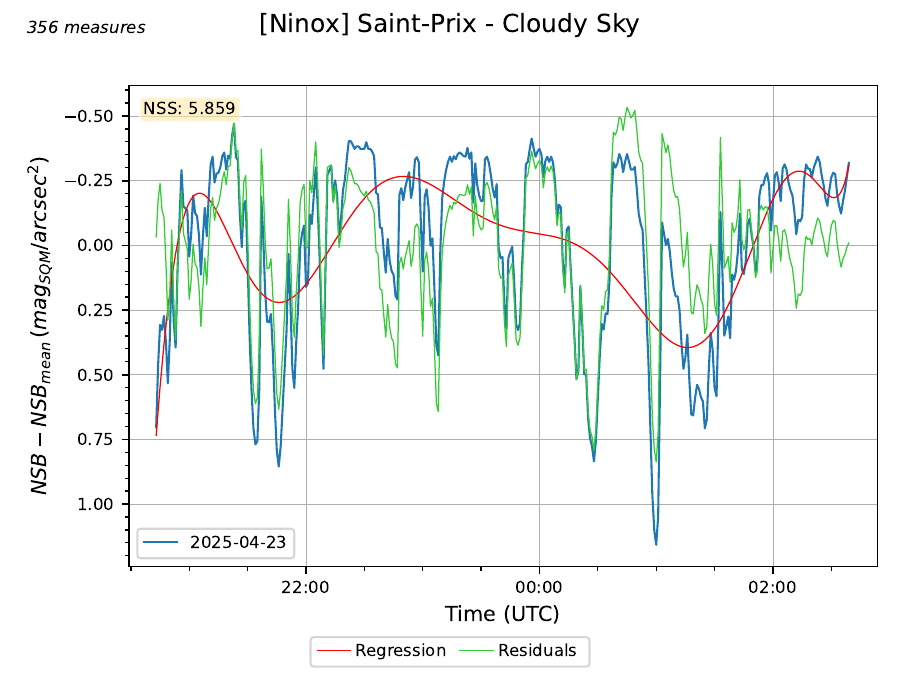}
\caption{This figure shows the NSB measurements (in blue), the corresponding 10$^{th}$-degree polynomial regression curve (in red) and the resulting residuals (in green). NSB data representative of a clear night is shown in the \textbf{top panel}, for cloudy night in the \textbf{bottom panel}. For convenience, the mean NSB of the night has been subtracted to the NSB values in the diagram. 
The NSS value is shown in the top left corner of each panel, it is significantly larger for cloudy nights \label{fig:nss}}
\end{figure}

\section{Additional light pollution parameter maps}
\label{app:maps}
\begin{figure}
\centering
\includegraphics[width=\textwidth]{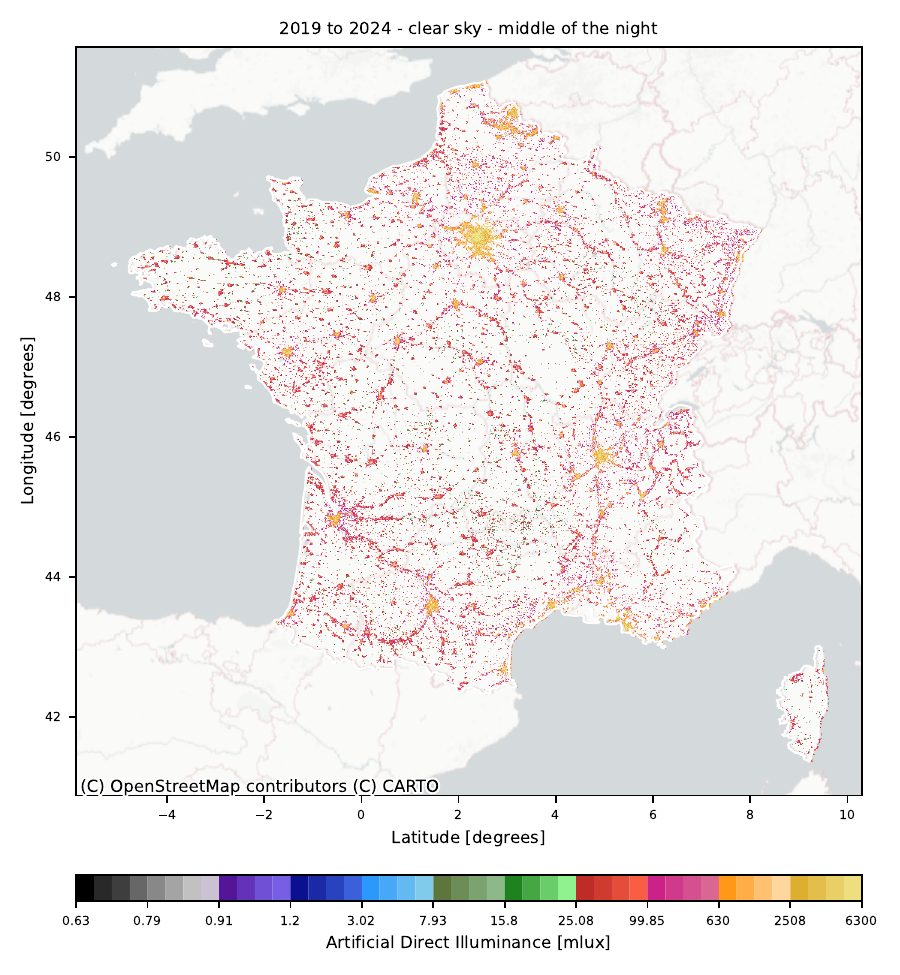}
\caption{Average direct ground illuminance for France in the middle of the night between 2019 and 2024 for clear sky conditions.}
\end{figure}   

\begin{figure}
\centering
\includegraphics[width=\textwidth]{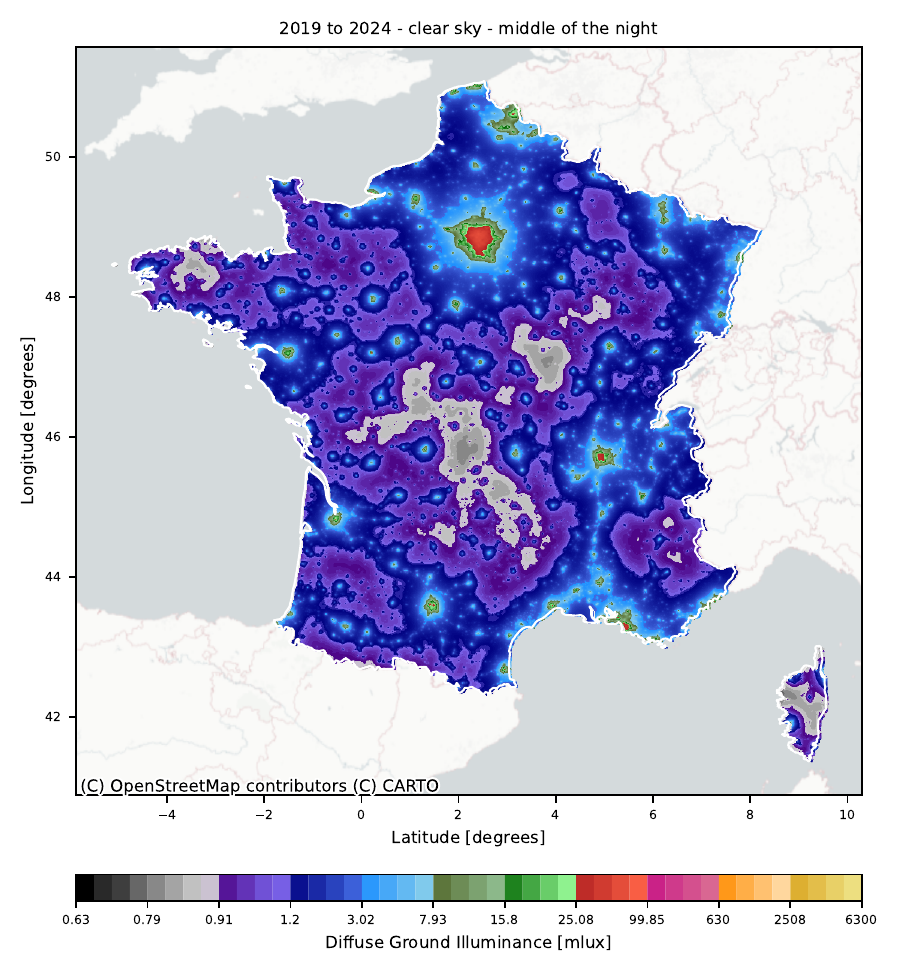}
\caption{Average diffuse ground illuminance for France in the middle of the night between 2019 and 2024. The latter is composed of the illuminance due to skyglow and the natural dark sky. The simulation was done for clear sky conditions.}
\end{figure}   

\begin{figure}
\centering
\includegraphics[width=\textwidth]{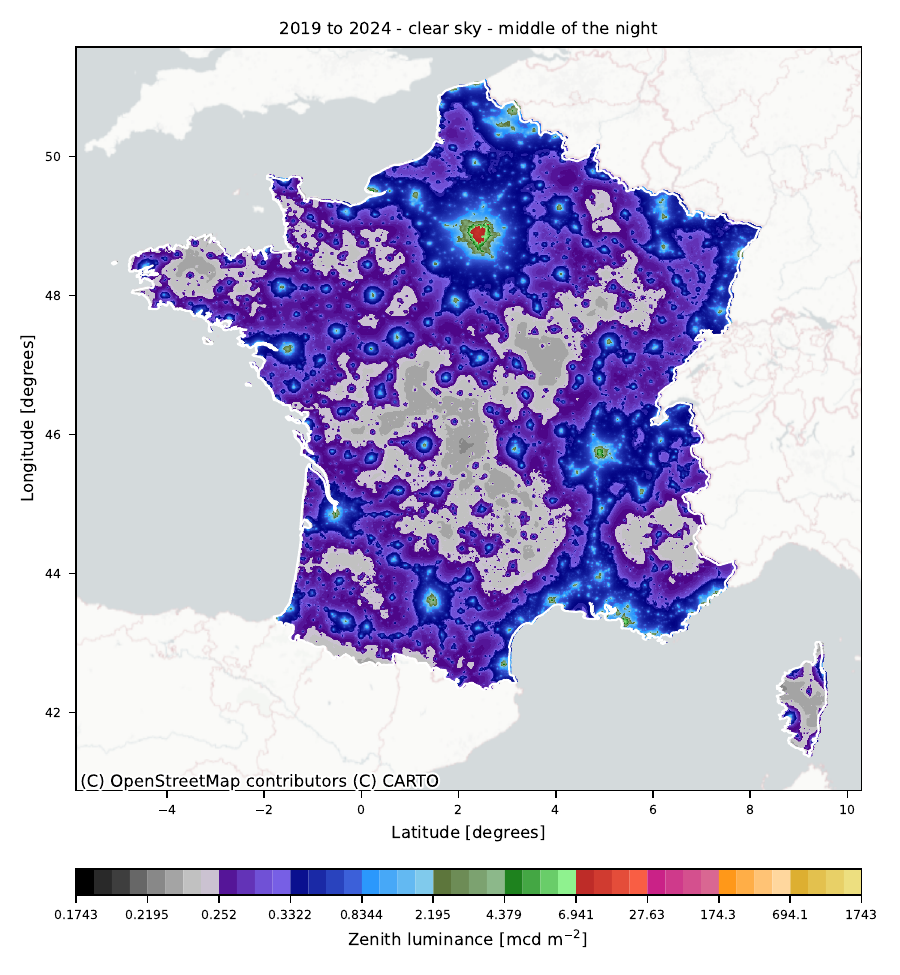}
\caption{Average total zenith luminance for France in the middle of the night between 2019 and 2024 for clear sky conditions. \label{fig:tzl}}
\end{figure}

\bibliographystyle{elsarticle-num}

\begin{thebibliography}{}
\expandafter\ifx\csname url\endcsname\relax
  \def\url#1{\texttt{#1}}\fi
\expandafter\ifx\csname urlprefix\endcsname\relax\def\urlprefix{URL }\fi
\expandafter\ifx\csname href\endcsname\relax
  \def\href#1#2{#2} \def\path#1{#1}\fi

\end{thebibliography}


\begin{thebibliography}{10}
	\expandafter\ifx\csname url\endcsname\relax
	\def\url#1{\texttt{#1}}\fi
	\expandafter\ifx\csname urlprefix\endcsname\relax\def\urlprefix{URL }\fi
	\expandafter\ifx\csname href\endcsname\relax
	\def\href#1#2{#2} \def\path#1{#1}\fi
	
	\bibitem{WANG2023121321}
	T.~Wang, N.~Kaida, K.~Kaida, \href{\tiny
		https://www.sciencedirect.com/science/article/pii/S0269749123003238}{Effects
		of outdoor artificial light at night on human health and behavior: A
		literature review}, Environmental Pollution 323 (2023) 121321.
	\newblock \href {https://doi.org/https://doi.org/10.1016/j.envpol.2023.121321}
	{\path{doi:https://doi.org/10.1016/j.envpol.2023.121321}}.
	\newline\urlprefix\url{\tiny
		https://www.sciencedirect.com/science/article/pii/S0269749123003238}
	
	\bibitem{Ma2024}
	S.~Ma, Y.~Alsabawi, H.~El-Serag, A.~Thrift, Exposure to light at night and risk
	of cancer: A systematic review, meta-analysis, and data synthesis, Cancers 16
	(2024) 2653.
	\newblock \href {https://doi.org/10.3390/cancers16152653}
	{\path{doi:10.3390/cancers16152653}}.
	
	\bibitem{OWENS2020108259}
	A.~C. Owens, P.~Cochard, J.~Durrant, B.~Farnworth, E.~K. Perkin, B.~Seymoure,
	\href{\tiny
		https://www.sciencedirect.com/science/article/pii/S0006320719307797}{Light
		pollution is a driver of insect declines}, Biological Conservation 241 (2020)
	108259.
	\newblock \href {https://doi.org/https://doi.org/10.1016/j.biocon.2019.108259}
	{\path{doi:https://doi.org/10.1016/j.biocon.2019.108259}}.
	\newline\urlprefix\url{\tiny
		https://www.sciencedirect.com/science/article/pii/S0006320719307797}
	
	\bibitem{Lao2019}
	S.~Lao, B.~Robertson, A.~Anderson, R.~Blair, J.~Eckles, R.~Turner, S.~Loss,
	Y.~Li, The influence of artificial light at night and polarized light on
	bird-building collisions, Biological Conservation 241 (2019) 108358.
	\newblock \href {https://doi.org/10.1016/j.biocon.2019.108358}
	{\path{doi:10.1016/j.biocon.2019.108358}}.
	
	\bibitem{Sanders2020}
	D.~Sanders, E.~Frago, R.~Kehoe, C.~Patterson, K.~Gaston, A meta-analysis of
	biological impacts of artificial light at night, Nature Ecology \& Evolution
	5 (2021) 1--8.
	\newblock \href {https://doi.org/10.1038/s41559-020-01322-x}
	{\path{doi:10.1038/s41559-020-01322-x}}.
	
	\bibitem{su11226400}
	M.~Grubisic, A.~Haim, P.~Bhusal, D.~M. Dominoni, K.~M.~A. Gabriel, A.~Jechow,
	F.~Kupprat, A.~Lerner, P.~Marchant, W.~Riley, K.~Stebelova, R.~H.~A. van
	Grunsven, M.~Zeman, A.~E. Zubidat, F.~Hölker, \href{\tiny
		https://www.mdpi.com/2071-1050/11/22/6400}{Light pollution, circadian
		photoreception, and melatonin in vertebrates}, Sustainability 11~(22) (2019).
	\newblock \href {https://doi.org/10.3390/su11226400}
	{\path{doi:10.3390/su11226400}}.
	\newline\urlprefix\url{\tiny https://www.mdpi.com/2071-1050/11/22/6400}
	
	\bibitem{Horton2023}
	K.~Horton, J.~Buler, S.~Anderson, C.~Burt, A.~Collins, A.~Dokter, F.~Guo,
	D.~Sheldon, M.~Tomaszewska, G.~Henebry, Artificial light at night is a top
	predictor of bird migration stopover density, Nature Communications 14 (12
	2023).
	\newblock \href {https://doi.org/10.1038/s41467-023-43046-z}
	{\path{doi:10.1038/s41467-023-43046-z}}.
	
	\bibitem{WangL2025}
	L.~Wang, L.~Meng, A.~Richardson, F.~Hölker, H.~Li, J.~Mao, T.~Longcore,
	J.~Xia, D.~She, Artificial light at night outweighs temperature in
	lengthening urban growing seasons, Nature Cities (06 2025).
	\newblock \href {https://doi.org/10.1038/s44284-025-00258-2}
	{\path{doi:10.1038/s44284-025-00258-2}}.
	
	\bibitem{10.1093/pnasnexus/pgac046}
	L.~Meng, Y.~Zhou, M.~O. Román, E.~C. Stokes, Z.~Wang, G.~R. Asrar, J.~Mao,
	A.~D. Richardson, L.~Gu, Y.~Wang, \href{\tiny
		https://doi.org/10.1093/pnasnexus/pgac046}{Artificial light at night: an
		underappreciated effect on phenology of deciduous woody plants}, PNAS Nexus
	1~(2) (2022) pgac046.
	\newblock \href
	{http://arxiv.org/abs/https://academic.oup.com/pnasnexus/article-pdf/1/2/pgac046/47087091/pgac046.pdf}
	{\path{arXiv:https://academic.oup.com/pnasnexus/article-pdf/1/2/pgac046/47087091/pgac046.pdf}},
	\href {https://doi.org/10.1093/pnasnexus/pgac046}
	{\path{doi:10.1093/pnasnexus/pgac046}}.
	\newline\urlprefix\url{\tiny https://doi.org/10.1093/pnasnexus/pgac046}
	
	\bibitem{barentine_2024_11431447}
	J.~Barentine, \href{\tiny https://doi.org/10.5281/zenodo.11431447}{Artificial
		light at night: State of the science 2024} (Jun. 2024).
	\newblock \href {https://doi.org/10.5281/zenodo.11431447}
	{\path{doi:10.5281/zenodo.11431447}}.
	\newline\urlprefix\url{\tiny https://doi.org/10.5281/zenodo.11431447}
	
	\bibitem{2022A&ARv..30....1G}
	R.~F. {Green}, C.~B. {Luginbuhl}, R.~J. {Wainscoat}, D.~{Duriscoe}, {The
		growing threat of light pollution to ground-based observatories}, A\&A~Rev.
	30~(1) (2022) 1.
	\newblock \href {https://doi.org/10.1007/s00159-021-00138-3}
	{\path{doi:10.1007/s00159-021-00138-3}}.
	
	\bibitem{2017SciA....3E1528K}
	C.~C.~M. {Kyba}, T.~{Kuester}, A.~{S{\'a}nchez de Miguel}, K.~{Baugh},
	A.~{Jechow}, F.~{H{\"o}lker}, J.~{Bennie}, C.~D. {Elvidge}, K.~J. {Gaston},
	L.~{Guanter}, {Artificially lit surface of Earth at night increasing in
		radiance and extent}, Science Advances 3~(11) (2017) e1701528.
	\newblock \href {https://doi.org/10.1126/sciadv.1701528}
	{\path{doi:10.1126/sciadv.1701528}}.
	
	\bibitem{2025JQSRT.33509378B}
	S.~{Bar{\'a}}, J.~J. {Castro-Torres}, {Diverging evolution of light pollution
		indicators: Can the globe at night and VIIRS-DNB measurements be
		reconciled?}, J.~Quant.~Spec.~Radiat.~Transf. 335 (2025) 109378.
	\newblock \href {http://arxiv.org/abs/2409.13111} {\path{arXiv:2409.13111}},
	\href {https://doi.org/10.1016/j.jqsrt.2025.109378}
	{\path{doi:10.1016/j.jqsrt.2025.109378}}.
	
	\bibitem{2023Sci...379..265K}
	C.~C.~M. {Kyba}, Y.~{\"O}. {Alt{\i}nta{\c{s}}}, C.~E. {Walker}, M.~{Newhouse},
	{Citizen scientists report global rapid reductions in the visibility of stars
		from 2011 to 2022}, Science 379~(6629) (2023) 265--268.
	\newblock \href {https://doi.org/10.1126/science.abq7781}
	{\path{doi:10.1126/science.abq7781}}.
	
	\bibitem{2000MNRAS.318..641C}
	P.~{Cinzano}, F.~{Falchi}, C.~D. {Elvidge}, K.~E. {Baugh}, {The artificial
		night sky brightness mapped from DMSP satellite Operational Linescan System
		measurements}, MNRAS 318~(3) (2000) 641--657.
	\newblock \href {http://arxiv.org/abs/astro-ph/0003412}
	{\path{arXiv:astro-ph/0003412}}, \href
	{https://doi.org/10.1046/j.1365-8711.2000.03562.x}
	{\path{doi:10.1046/j.1365-8711.2000.03562.x}}.
	
	\bibitem{2001MNRAS.328..689C}
	P.~{Cinzano}, F.~{Falchi}, C.~D. {Elvidge}, {The first World Atlas of the
		artificial night sky brightness}, MNRAS 328~(3) (2001) 689--707.
	\newblock \href {http://arxiv.org/abs/astro-ph/0108052}
	{\path{arXiv:astro-ph/0108052}}, \href
	{https://doi.org/10.1046/j.1365-8711.2001.04882.x}
	{\path{doi:10.1046/j.1365-8711.2001.04882.x}}.
	
	\bibitem{2012MNRAS.427.3337C}
	P.~{Cinzano}, F.~{Falchi}, {The propagation of light pollution in the
		atmosphere}, MNRAS 427~(4) (2012) 3337--3357.
	\newblock \href {http://arxiv.org/abs/1209.2031} {\path{arXiv:1209.2031}},
	\href {https://doi.org/10.1111/j.1365-2966.2012.21884.x}
	{\path{doi:10.1111/j.1365-2966.2012.21884.x}}.
	
	\bibitem{2016SciA....2E0377F}
	F.~{Falchi}, P.~{Cinzano}, D.~{Duriscoe}, C.~C.~M. {Kyba}, C.~D. {Elvidge},
	K.~{Baugh}, B.~A. {Portnov}, N.~A. {Rybnikova}, R.~{Furgoni}, {The new world
		atlas of artificial night sky brightness}, Science Advances 2~(6) (2016)
	e1600377--e1600377.
	\newblock \href {http://arxiv.org/abs/1609.01041} {\path{arXiv:1609.01041}},
	\href {https://doi.org/10.1126/sciadv.1600377}
	{\path{doi:10.1126/sciadv.1600377}}.
	
	\bibitem{2020MNRAS.497.2501A}
	M.~{Aub{\'e}}, A.~{Simoneau}, C.~{Mu{\~n}oz-Tu{\~n}{\'o}n},
	J.~{D{\'\i}az-Castro}, M.~{Serra-Ricart}, {Restoring the night sky darkness
		at Observatorio del Teide: First application of the model Illumina version
		2}, MNRAS 497~(3) (2020) 2501--2516.
	\newblock \href {http://arxiv.org/abs/2005.14160} {\path{arXiv:2005.14160}},
	\href {https://doi.org/10.1093/mnras/staa2113}
	{\path{doi:10.1093/mnras/staa2113}}.
	
	\bibitem{Dreyer2025-hd}
	D.~Dreyer, A.~Adden, H.~Chen, B.~Frost, H.~Mouritsen, J.~Xu, K.~Green,
	M.~Whitehouse, J.~Chahl, J.~Wallace, G.~Hu, J.~Foster, S.~Heinze, E.~Warrant,
	Bogong moths use a stellar compass for long-distance navigation at night,
	Nature 643~(8073) (2025) 994--1000.
	
	\bibitem{2015MNRAS.446.2895K}
	M.~{Kocifaj}, T.~{Posch}, H.~A. {Solano Lamphar}, {On the relation between
		zenith sky brightness and horizontal illuminance}, MNRAS 446~(3) (2015)
	2895--2901.
	\newblock \href {https://doi.org/10.1093/mnras/stu2265}
	{\path{doi:10.1093/mnras/stu2265}}.
	
	\bibitem{2025MNRAS.542L.154K}
	M.~{Kocifaj}, F.~{Falchi}, {Generalizing the world atlas of artificial night
		sky brightness to encompass a wide range of sky states}, MNRAS 542~(1)
	(2025) L154--L157.
	\newblock \href {https://doi.org/10.1093/mnrasl/slaf083}
	{\path{doi:10.1093/mnrasl/slaf083}}.
	
	\bibitem{Cerema2020}
	Cerema, {Décryptage : l'arrêté ministériel "nuisances lumineuses" -
		Contexte}, \url{ \tiny
		https://www.cerema.fr/fr/actualites/decryptage-arrete-ministeriel-nuisances-lumineuses-contexte},
	[Online; accessed August-2025] (2020).
	
	\bibitem{Cerema2025}
	Cerema, {Extinction de l'éclairage public : une étude sur les pratiques des
		collectivités}, \url{ \tiny
		https://www.cerema.fr/fr/actualites/extinction-eclairage-public-etude-pratiques-collectivites},
	[Online; accessed August-2025] (2025).
	
	\bibitem{2020MNRAS.493.2429B}
	S.~{Bar{\'a}}, M.~{Aub{\'e}}, J.~{Barentine}, J.~{Zamorano}, {Magnitude to
		luminance conversions and visual brightness of the night sky}, MNRAS 493~(2)
	(2020) 2429--2437.
	\newblock \href {http://arxiv.org/abs/2002.01494} {\path{arXiv:2002.01494}},
	\href {https://doi.org/10.1093/mnras/staa323}
	{\path{doi:10.1093/mnras/staa323}}.
	
	\bibitem{1986PASP...98..364G}
	R.~H. {Garstang}, {Model for artificial night-sky illumination.}, PASP 98
	(1986) 364--375.
	\newblock \href {https://doi.org/10.1086/131768} {\path{doi:10.1086/131768}}.
	
	\bibitem{1989PASP..101..306G}
	R.~H. {Garstang}, {Night Sky Brightness at Observatories and Sites}, PASP 101
	(1989) 306.
	\newblock \href {https://doi.org/10.1086/132436} {\path{doi:10.1086/132436}}.
	
	\bibitem{doi:10.1177/1477153513506729}
	D.~Duriscoe, C.~Luginbuhl, C.~Elvidge, \href{\tiny
		https://doi.org/10.1177/1477153513506729}{The relation of outdoor lighting
		characteristics to sky glow from distant cities}, Lighting Research \&
	Technology 46~(1) (2014) 35--49.
	\newblock \href {https://doi.org/10.1177/1477153513506729}
	{\path{doi:10.1177/1477153513506729}}.
	\newline\urlprefix\url{\tiny https://doi.org/10.1177/1477153513506729}
	
	\bibitem{2014RemS....611915C}
	C.~{Cao}, Y.~{Bai}, {Quantitative Analysis of VIIRS DNB Nightlight Point Source
		for Light Power Estimation and Stability Monitoring}, Remote Sensing 6~(12)
	(2014) 11915--11935.
	\newblock \href {https://doi.org/10.3390/rs61211915}
	{\path{doi:10.3390/rs61211915}}.
	
	\bibitem{2000MmSAI..71...93C}
	P.~{Cinzano}, {The propagation of light pollution in diffusely urbanised
		areas}, Mem.~Soc.~Astron.~Italiana 71 (2000) 93.
	
	\bibitem{2014MNRAS.443.3665K}
	M.~{Kocifaj}, H.~A. {Solano Lamphar}, {Quantitative analysis of night skyglow
		amplification under cloudy conditions}, MNRAS 443~(4) (2014) 3665--3674.
	\newblock \href {https://doi.org/10.1093/mnras/stu1301}
	{\path{doi:10.1093/mnras/stu1301}}.
	
	\bibitem{2016JQSRT.181...11A}
	M.~{Aub{\'e}}, M.~{Kocifaj}, J.~{Zamorano}, H.~A. {Solano Lamphar}, A.~{Sanchez
		de Miguel}, {The spectral amplification effect of clouds to the night sky
		radiance in Madrid}, J.~Quant.~Spec.~Radiat.~Transf. 181 (2016) 11--23.
	\newblock \href {https://doi.org/10.1016/j.jqsrt.2016.01.032}
	{\path{doi:10.1016/j.jqsrt.2016.01.032}}.
	
	\bibitem{2017NatSR...7.6741J}
	A.~{Jechow}, Z.~{Koll{\'a}th}, S.~J. {Ribas}, H.~{Spoelstra}, F.~{H{\"o}lker},
	C.~C.~M. {Kyba}, {Imaging and mapping the impact of clouds on skyglow with
		all-sky photometry}, Scientific Reports 7 (2017) 6741.
	\newblock \href {http://arxiv.org/abs/1705.04968} {\path{arXiv:1705.04968}},
	\href {https://doi.org/10.1038/s41598-017-06998-z}
	{\path{doi:10.1038/s41598-017-06998-z}}.
	
	\bibitem{1991PASP..103.1109G}
	R.~H. {Garstang}, {Dust and Light Pollution}, PASP 103 (1991) 1109.
	\newblock \href {https://doi.org/10.1086/132933} {\path{doi:10.1086/132933}}.
	
	\bibitem{2023JEnvM.33517534W}
	S.~{Wallner}, M.~{Kocifaj}, {Aerosol impact on light pollution in cities and
		their environment}, Journal of Environmental Management 335 (2023) 17534.
	\newblock \href {https://doi.org/10.1016/j.jenvman.2023.117534}
	{\path{doi:10.1016/j.jenvman.2023.117534}}.
	
	\bibitem{2023MNRAS.523.2678K}
	M.~{Kocifaj}, F.~{Kundracik}, J.~{Barentine}, {Aerosol parameters for night sky
		brightness modelling estimated from daytime sky images}, MNRAS 523~(2)
	(2023) 2678--2683.
	\newblock \href {http://arxiv.org/abs/2306.06750} {\path{arXiv:2306.06750}},
	\href {https://doi.org/10.1093/mnras/stad1570}
	{\path{doi:10.1093/mnras/stad1570}}.
	
	\bibitem{2024MNRAS.533.2356K}
	M.~{Kocifaj}, J.~{Petr{\v{z}}ala}, I.~{Medve{\v{d}}}, {Skyglow from
		ground-reflected radiation: model improvements},  533~(2) (2024)
	2356--2363.
	\newblock \href {https://doi.org/10.1093/mnras/stae1992}
	{\path{doi:10.1093/mnras/stae1992}}.
	
	\bibitem{ROMAN2018113}
	M.~O. Román, Z.~Wang, Q.~Sun, V.~Kalb, S.~D. Miller, A.~Molthan, L.~Schultz,
	J.~Bell, E.~C. Stokes, B.~Pandey, K.~C. Seto, D.~Hall, T.~Oda, R.~E. Wolfe,
	G.~Lin, N.~Golpayegani, S.~Devadiga, C.~Davidson, S.~Sarkar, C.~Praderas,
	J.~Schmaltz, R.~Boller, J.~Stevens, O.~M. {Ramos González}, E.~Padilla,
	J.~Alonso, Y.~Detrés, R.~Armstrong, I.~Miranda, Y.~Conte, N.~Marrero,
	K.~MacManus, T.~Esch, E.~J. Masuoka, \href{\tiny
		https://www.sciencedirect.com/science/article/pii/S003442571830110X}{Nasa's
		black marble nighttime lights product suite}, Remote Sensing of Environment
	210 (2018) 113--143.
	\newblock \href {https://doi.org/https://doi.org/10.1016/j.rse.2018.03.017}
	{\path{doi:https://doi.org/10.1016/j.rse.2018.03.017}}.
	\newline\urlprefix\url{\tiny
		https://www.sciencedirect.com/science/article/pii/S003442571830110X}
	
	\bibitem{BlackMarbleV2}
	Z.~Wang, M.~O. Román, R.~Shrestha, T.~Yao, V.~Kalb, {Black Marble User Guide
		(Collection 2.0)}, \url{\tiny
		https://viirsland.gsfc.nasa.gov/PDF/BlackMarbleUserGuide\_Collection2.0.pdf},
	[Online; accessed June-2025] (2024).
	
	\bibitem{2019RSEnv.23311357L}
	X.~{Li}, R.~{Ma}, Q.~{Zhang}, D.~{Li}, S.~{Liu}, T.~{He}, L.~{Zhao},
	{Anisotropic characteristic of artificial light at night - Systematic
		investigation with VIIRS DNB multi-temporal observations}, Remote Sensing of
	Environment 233 (2019) 111357.
	\newblock \href {https://doi.org/10.1016/j.rse.2019.111357}
	{\path{doi:10.1016/j.rse.2019.111357}}.
	
	\bibitem{sanchez2020}
	A.~Sanchez~de Miguel, C.~C.~M. Kyba, J.~Zamorano, J.~Gallego, K.~J. Gaston,
	\href{\tiny https://www.nature.com/articles/s41598-020-64673-2}{The nature of
		the diffuse light near cities detected in nighttime satellite imagery},
	Scientific Reports 10 (2020).
	\newblock \href {https://doi.org/10.1038/s41598-020-64673-2}
	{\path{doi:10.1038/s41598-020-64673-2}}.
	\newline\urlprefix\url{\tiny
		https://www.nature.com/articles/s41598-020-64673-2}
	
	\bibitem{sdgsat}
	H.~Guo, C.~Dou, H.~Chen, J.~Liu, B.~Fu, X.~Li, Z.~Zou, D.~Liang, Sdgsat-1: the
	world’s first scientific satellite for sustainable development goals,
	Science Bulletin 68 (12 2022).
	\newblock \href {https://doi.org/10.1016/j.scib.2022.12.014}
	{\path{doi:10.1016/j.scib.2022.12.014}}.
	
	\bibitem{2024ITGRS..6270572Y}
	L.~{Yan}, Y.~{Hu}, C.~{Dou}, X.-M. {Li}, {Radiometric Calibration of SDGSAT-1
		Nighttime Light Payload}, IEEE Transactions on Geoscience and Remote Sensing
	62 (2024) 3370572.
	\newblock \href {https://doi.org/10.1109/TGRS.2024.3370572}
	{\path{doi:10.1109/TGRS.2024.3370572}}.
	
	\bibitem{DEVERCHERE2022}
	P.~Deverchere, S.~Vauclair, G.~Bosch, S.~Moulherat, J.~Cornuau, Towards an
	absolute light pollution indicator, Scientific Reports 12 (10 2022).
	\newblock \href {https://doi.org/10.1038/s41598-022-21460-5}
	{\path{doi:10.1038/s41598-022-21460-5}}.
	
	\bibitem{SkyGlowWeb}
	M.~Kocifaj, {SkyGlow software documentation}, \url{ \tiny
		https://skyglow.sav.sk/\#simulator}, [Online; accessed June-2025] (2025).
	
	\bibitem{DiAntonio2023}
	L.~Di~Antonio, C.~Di~Biagio, G.~Foret, P.~Formenti, G.~Siour, J.-F. Doussin,
	M.~Beekmann, Aerosol optical depth climatology from the high-resolution maiac
	product over europe: differences between major european cities and their
	surrounding environments, Atmospheric Chemistry and Physics 23 (2023)
	12455--12475.
	\newblock \href {https://doi.org/10.5194/acp-23-12455-2023}
	{\path{doi:10.5194/acp-23-12455-2023}}.
	
	\bibitem{philippe_deverchere_2025_17098451}
	P.~Deverchère, \href{https://doi.org/10.5281/zenodo.17098451}{Ninox csb
		database} (Sep. 2025).
	\newblock \href {https://doi.org/10.5281/zenodo.17098451}
	{\path{doi:10.5281/zenodo.17098451}}.
	\newline\urlprefix\url{https://doi.org/10.5281/zenodo.17098451}
	
	\bibitem{2021AJ....162...25A}
	M.~R. {Alarcon}, M.~{Serra-Ricart}, S.~{Lemes-Perera}, M.~{Mallorqu{\'\i}n},
	{Natural Night Sky Brightness during Solar Minimum}, AJ 162~(1) (2021) 25.
	\newblock \href {http://arxiv.org/abs/2105.01066} {\path{arXiv:2105.01066}},
	\href {https://doi.org/10.3847/1538-3881/abfdaa}
	{\path{doi:10.3847/1538-3881/abfdaa}}.
	
	\bibitem{doi:10.1126/sciadv.abl6891}
	A.~S. de~Miguel, J.~Bennie, E.~Rosenfeld, S.~Dzurjak, K.~J. Gaston,
	\href{https://www.science.org/doi/abs/10.1126/sciadv.abl6891}{Environmental
		risks from artificial nighttime lighting widespread and increasing across
		europe}, Science Advances 8~(37) (2022) eabl6891.
	\newblock \href
	{http://arxiv.org/abs/https://www.science.org/doi/pdf/10.1126/sciadv.abl6891}
	{\path{arXiv:https://www.science.org/doi/pdf/10.1126/sciadv.abl6891}}, \href
	{https://doi.org/10.1126/sciadv.abl6891} {\path{doi:10.1126/sciadv.abl6891}}.
	\newline\urlprefix\url{https://www.science.org/doi/abs/10.1126/sciadv.abl6891}
	
	\bibitem{Bortle2001}
	J.~E. Bortle, Gauging light pollution: The bortle dark-sky scale, Sky \&
	Telescope February (2 2001).
	
	\bibitem{Baray2019}
	J.~Baray, A.~Bah, P.~Cacault, K.~Sellegri, J.-M. Pichon, L.~Deguillaume,
	N.~Montoux, V.~Noel, G.~Seze, F.~Gabarrot, G.~Payen, V.~Duflot, Cloud
	occurrence frequency at puy de dôme (france) deduced from an automatic
	camera image analysis: Method, validation, and comparisons with larger scale
	parameters, Atmosphere 10 (12 2019).
	\newblock \href {https://doi.org/10.3390/atmos10120808}
	{\path{doi:10.3390/atmos10120808}}.
	
	\bibitem{TESSManual}
	L.~García, J.~Zamorano, C.~Tapia, {TESS Photometer Manual}, \url{ \tiny
		https://guaix.fis.ucm.es/sites/guaix.fis.ucm.es.tess/files/documents/2020-06-v3\_TESS-W\_manual\_english.pdf},
	[Online; accessed August-2025] (2019).
	
\end{thebibliography}

\end{document}